%% file: pointing.tex
\documentclass[letterpaper]{ws-jai}

\usepackage{amsmath}
\usepackage{gensymb}
\usepackage{lineno}
\usepackage{float}
\usepackage[caption = false]{subfig}
\usepackage{url}
\usepackage{threeparttable}
\usepackage{xcolor}
\definecolor{byzantine}{rgb}{0.74, 0.2, 0.64}
\usepackage{multicol}
\usepackage[
	colorlinks,
	citecolor=byzantine,
	linkcolor=byzantine,
	urlcolor=byzantine
]{hyperref}
\usepackage{orcidlink}

\makeatletter
\let\trimmarks\relax
\let\wsdraftnote\relax
\makeatother

\DeclareMathOperator{\az}{\mathit{A}}
\DeclareMathOperator{\Daz}{\Delta\az}
\DeclareMathOperator{\el}{\mathit{E}}
\DeclareMathOperator{\Del}{\Delta\el}
\def \aztext {azimuth}
\def \eltext {elevation}

\def \Eltext {Elevation}
\def \xeltext {cross-elevation}
\def \Xeltext {Cross-elevation}
\def \azmath {$\az$}
\def \elmath {$\el$}

\DeclareMathOperator{\azero}{\phi_0}
\DeclareMathOperator{\aone}{\phi_1}
\DeclareMathOperator{\atwo}{\tau_{ax}}
\DeclareMathOperator{\athree}{\tau_{ay}}
\DeclareMathOperator{\afour}{\tau_{e}}
\DeclareMathOperator{\afive}{\kappa_x}
\DeclareMathOperator{\asix}{\kappa_y}
\DeclareMathOperator{\aseven}{\overline{\el}}
\DeclareMathOperator{\azzero}{\overline{\az}}

\newcommand{\SIN}[1]{\sin{ \left( #1 \right) }}
\newcommand{\COS}[1]{\cos{ \left( #1 \right) }}

\newcommand{\ASIN}[1]{\sin^{-1}{ \left( #1 \right) }}
\newcommand{\ACOS}[1]{\cos^{-1}{ \left( #1 \right) }}
\newcommand{\ATAN}[1]{\tan^{-1}{ \left( #1 \right) }}

\newcommand\arcmin{\mbox{$^\prime$}}%
\newcommand\arcsec{\mbox{$^{\prime\prime}$}}%
\newcommand\farcm{\mbox{$.\mkern-4mu^\prime$}}%
\newcommand\farcs{\mbox{$.\!\!^{\prime\prime}$}}%

\begin{document}

\pdfpagewidth=8.5in
\pdfpageheight=11in

\catchline{}{}{}{}{} 

\markboth{P. M. Chichura et al.}{Pointing Accuracy Improvements for the SPT with ML}

\title{Pointing Accuracy Improvements for the South Pole Telescope with Machine Learning}

\input{authors}
\corres{$^*$Corresponding author.}
\maketitle

\begin{history}
\received{January 3, 2025};
\accepted{April 23, 2025}
\end{history}

\begin{abstract}
We present improvements to the pointing accuracy of the South Pole Telescope (SPT) using machine learning.
The ability of the SPT to point accurately at the sky is limited by its structural imperfections, which are impacted by the extreme weather at the South Pole.
Pointing accuracy is particularly important during SPT participation in observing campaigns with the Event Horizon Telescope (EHT), which requires stricter accuracy than typical observations with the SPT.
We compile a training dataset of historical observations of astronomical sources made with the SPT-3G and EHT receivers on the SPT.
We train two XGBoost models to learn a mapping from current weather conditions to two telescope drive control arguments---one which corrects for errors in azimuth and the other for errors in elevation.
Our trained models achieve root mean squared errors on withheld test data of 2\farcs14 in \xeltext{} and 3\farcs57 in \eltext{}, well below our goal of 5\arcsec{} along each axis.
We deploy our models on the telescope control system and perform further \textit{in situ} test observations during the EHT observing campaign in 2024 April.
Our models result in significantly improved pointing accuracy: for sources within the range of input variables where the models are best trained, average combined pointing error improved 33\%, from 15\farcs9 to 10\farcs6.
These improvements, while significant, fall shy of our ultimate goal, but they serve as a proof of concept for the development of future models.
Planned upgrades to the EHT receiver on the SPT will necessitate even stricter pointing accuracy which will be achievable with our methods.

\end{abstract}


\begin{multicols}{2}

\section{Introduction}
 

To observe, telescopes must be able to point towards a desired location on the sky and to do so accurately and repeatedly.
The process of orienting the telescope aperture on the sky is often simply called ``pointing.''
The required pointing accuracy is determined by the size of the telescope ``beam,'' i.e., the angular dependence of the response of the telescope receiver to incoming radiation from the sky.
Telescopes have finite resolution; any structures on the sky smaller than the beam will appear blurred by the shape of the beam, reflecting a convolution of the true sky brightness with the telescope beam.
When the telescope is pointed directly toward a source in the sky, the response of the receiver to that source is largest, and as the telescope pointing is offset from the source, the response tapers off.
Errors in pointing, often called ``pointing offsets,'' occur when the telescope is not aligned as instructed.
Unwanted pointing offsets affect the quality of data, so astronomers will typically require that telescope control systems are reliable, with pointing errors less than some fraction of the beam.

Pointing errors can be caused by many physical processes.
These processes include but are not limited to deformations and tilts in the support structure, flexure of the mirrors or receiver, and misalignment of optical axes (``collimation'' errors).
Most of these processes can be accounted for by analytically modeling their effects on pointing.
Such ``pointing models'' are typically a series of spherical trigonometry equations that relate an instructed telescope orientation to an actual sky coordinate after accounting for any deformations in the telescope structure.
The magnitude of each type of deformation is described by a handful of adjustable parameters in the pointing model.
For more details on pointing model theory, we refer readers to \citet{stumpff72} and \citet{ulich81}.
Determining the correct values of pointing model parameters is a critical task for any telescope.
Some examples of efforts to maintain pointing accuracy for telescopes include \citet{vonhoerner75} for the 140-ft telescope at Green Bank, \citet{greve96} for the IRAM 30-m telescope, \citet{baars99} for the Heinrich Hertz Telescope, \citet{gawronski05} for the Large Millimeter Telescope, and \citet{white22} for the Green Bank Telescope. 

\end{multicols}
\twocolumn

The South Pole Telescope (SPT) is located at the National Science Foundation's Amundsen-Scott South Pole research station, approximately 1 km from the geographic South Pole \cite{carlstrom11}.
The SPT 10-m primary mirror is optimized for observing at millimeter/sub-millimeter wavelengths.
Since its construction, the SPT has primarily been outfitted with a series of three cameras, the most recent of which is the SPT-3G camera \cite{sobrin22}.
The primary science goal of the SPT is to characterize the cosmic microwave background (CMB), light generated in the early universe nearly 14~billion years ago.
To achieve this goal, the SPT is used to observe a particular region of the sky repeatedly over many years to ensure low measurement uncertainty.

The SPT control system utilizes a pointing model with nine parameters, described later in Section~\ref{sec:pointing_model}.
The values of these parameters were determined by observing astronomical sources with known sky positions, recording the apparent positions, and later fitting the pointing model for parameter values which would have best mitigated the offsets between apparent and true positions.
In the standard observing mode, we hold these parameters constant from observation to observation, which is sufficiently accurate for our primary science goal.
The SPT, with its 10-m primary mirror, is capable of observing many bright point sources with high S/N in each CMB survey observation.
Any pointing errors in these observations can be corrected in post-processing by shifting the images so that the sources appear where expected.


However, there are some uses of the SPT for which pointing errors are more consequential.
The SPT is part of the Event Horizon Telescope (EHT), a very-long-baseline interferometry experiment consisting of telescopes across the world observing at 230~GHz and 345~GHz observing frequencies \cite{doeleman10}.
A primary science goal of the EHT is to study the immediate environment around black holes, and EHT observations have resulted in the first-ever images of black holes at the event horizon scale \cite{eht19a, eht22a}.
This goal requires extraordinarily high angular resolution.
The ultimate angular resolution of the EHT configuration is driven in part by the uniquely southern geographic location of the SPT, which provides several long baselines to member telescopes at other sites on Earth.
For about three weeks every year, the SPT takes part in scheduled EHT observing campaigns by switching observing modes to use a separate EHT receiver \cite{kim18b}, the first results from which are presented in \citet{kim18a}.
The EHT receiver consists of a single pixel, and EHT observations require that pixel to consistently point directly at the astronomical source of interest.
Because EHT observations do not create images of the sky like typical SPT observations, we cannot correct for pointing offsets after the fact.

The design requirements for the SPT specify 4\arcsec{} pointing accuracy, which is much smaller than the SPT-3G beam sizes of 1\farcm57, 1\farcm17, and 1\farcm04 at 95~GHz, 150~GHz, and 220~GHz, respectively \cite{sobrin22}.
The specifications were met when the telescope was first constructed in Texas in 2006 prior to deployment to Antarctica.
However, the specifications were not met at the South Pole, primarily because of large thermal gradients in the telescope structure, driven by cold outside temperatures and the need to keep the telescope bearings and drives warm.
Despite extraordinarily stable weather at the South Pole, especially when compared to sites of other comparable telescopes, temperatures typically vary around mean values of -$25\degree$C in the austral summer and -$60\degree$C in the austral winter \cite{lazzara12}.
Extreme minimum temperatures during most austral winters reach below -$73\degree$C.
These conditions, in conjunction with a heated control room at the telescope base, cause a large thermal gradient through the telescope support structure.
Telescopes in less extreme conditions, including the SPT while in Texas, do not encounter such severe thermal differences with their environments.
Thermal gradients induce thermal deformations in the telescope which change with the weather.
To correct for these varying deformations, we require dynamic pointing model parameters.
While assuming constant pointing model parameters is sufficient for typical SPT observations, such an assumption is not sufficient for EHT observations.
Currently, the SPT manages pointing errors during participation in EHT observations by performing corrective observations at some cost to observing efficiency, but planned upgrades to the EHT receiver will require even finer control of pointing offsets.

To meet the requirements for EHT observations, we must be able to correct for variations in the pointing model parameters caused by the changing weather.
We previously attempted to model these changes analytically based on physical intuition.
The telescope is outfitted with many sensors to monitor its current status: length sensors record important structural dimensions, thermometers record temperature throughout the instrument, and a nearby weather station monitors weather conditions like air temperature, wind speed, and wind direction \cite{carlstrom11}.
However, analytical modeling fails to capture much of the complexity present in the data.
We require more complex modeling for the thermal deformations of the SPT.

Other telescopes have encountered similar issues.
For instance, \citet{antebi98} presented efforts to account in real time for thermal deformations in the Large Millimeter Telescope.
\citet{mittag08} detailed a correlation between ambient temperature and pointing accuracy of the Hamburg Robotic Telescope.
\citet{dong18} measured and analyzed the thermal deformations of the Tianma Radio Telescope.
\citet{greve10} presented a detailed discussion of the thermal behavior of radio telescopes.
Recently, some researchers have made attempts at more complex modeling with machine learning (ML).
\citet{hou23} developed an ML-based model for estimating and predicting seeing at Dome A, Antarctica.
\citet{nyheim24} developed an ML-based pointing model for the Atacama Pathfinder EXperiment.

Here, we present our efforts to train two ML models which can correct for pointing errors of the SPT in real time.
We train the models to take information about a target source location, the current state of the telescope, and current weather conditions, then estimate pointing model parameters which yield accurate pointing.
We integrate these models into the telescope control system so that the system can adjust its pointing on demand.
We then evaluate the performance of our models with test observations.
We find that these ML models significantly improve the pointing accuracy of the SPT but require further development to fully reach our goal.

In Section~\ref{sec:current_pointing_methods}, we describe the current pointing methods used on the telescope for typical SPT and EHT observations.
In Section~\ref{sec:ml_training}, we present the methods used to train ML models to correct SPT pointing errors.
In Section~\ref{sec:results}, we describe deploying the trained ML models on the telescope and performing real-world test observations.
In Section~\ref{sec:discussion}, we discuss the results of our tests and the challenges that need to be overcome to further improve the pointing accuracy of the SPT.

\section{Current Pointing Methods}
\label{sec:current_pointing_methods}

\begin{figure*}[h]
	\centering
	\includegraphics[width=0.8\linewidth]{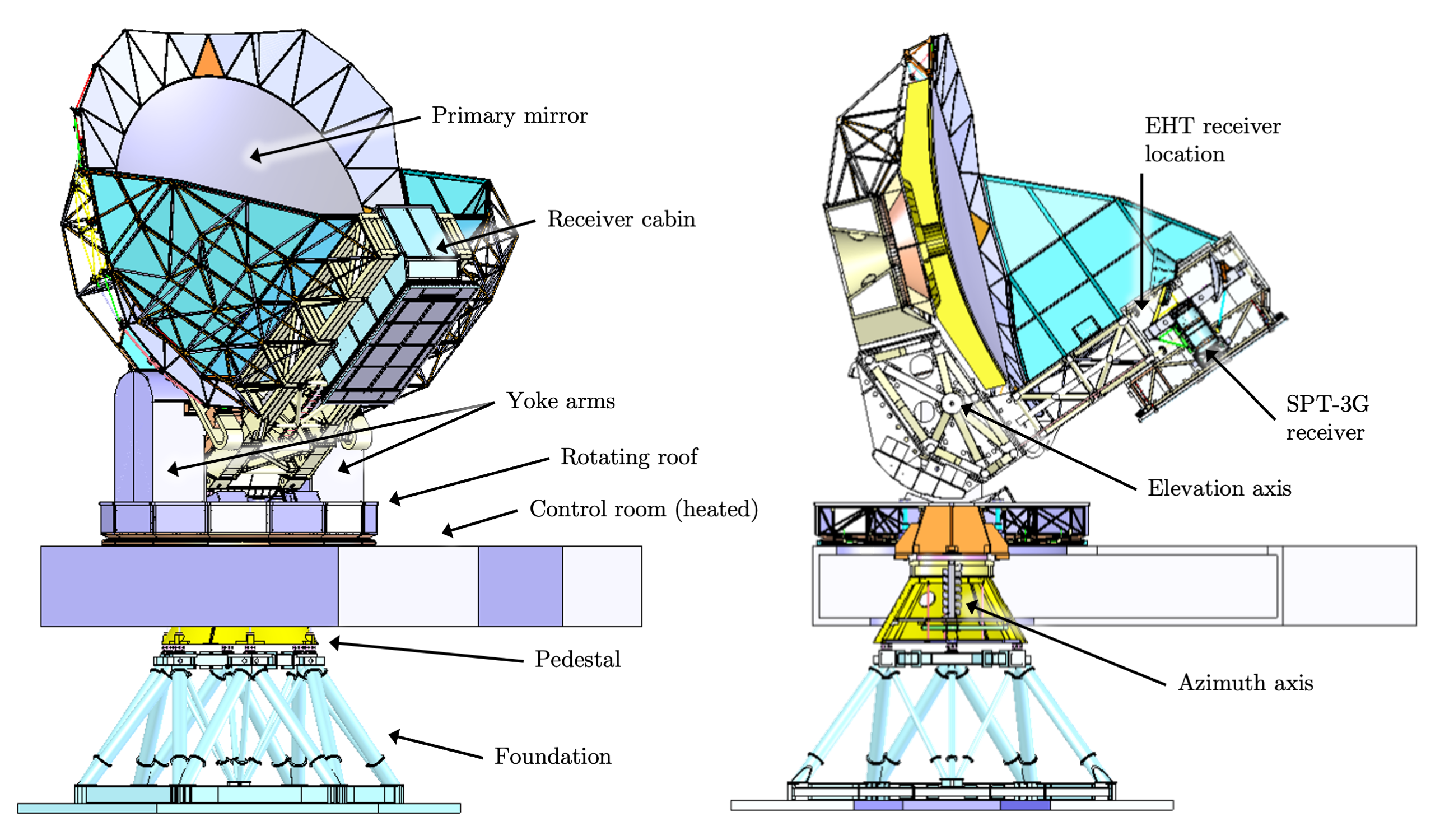}
	\caption{
		\small
		A rendering of the SPT from \textit{(left)} external and \textit{(right)} internal cross-section views.
		Parts of the telescope key to discussing pointing accuracy are labeled.
	}
	\label{fig:spt}
\end{figure*}

To observe some target source in the sky, we must point the telescope towards the location of that source.
The SPT is mounted on an altazimuth structure, which enables rotation about two perpendicular axes \cite{carlstrom11}.
Rotation about the vertical axis varies azimuth (\azmath), or angular direction along the horizon.
By definition, \aztext{} spans $0\degree \leq \az < 360\degree$, where $\az = 0\degree$ points northward\footnote{At the location of the SPT, azimuth $\az = 0\degree$ aligns with true north, which is different from grid north ($\az \approx 44\degree$), another coordinate system frequently used at the South Pole.} and increases in the clockwise direction.
Rotation about the horizontal axis varies elevation (\elmath), or angular direction above the horizon.
By definition, \eltext{} spans $-90\degree \leq \el \leq 90\degree$, where $\el = 0\degree$ points towards the horizon and increases upwards.
The SPT can be instructed to the location of any source on the sky with a pair of \aztext{} and \eltext{} angles.
Figure~\ref{fig:spt} shows the locations of the \aztext{} and \eltext{} axes along with other key telescope features discussed in this section.

Rotation about the mount axes is controlled by a set of motors and angle encoders \cite{carlstrom11}.
The motors rotate the bearings of the axes, and the angle encoders measure rotational position along those axes.
When a desired sky location is instructed, motors at the base of the telescope rotate the telescope in \aztext{} until the \aztext{} encoder measures the desired angle.
Similarly, two arms called the ``yoke arms'' support the primary mirror and receiver cabin, and motors in each arm rotate this structure in \eltext{} until \eltext{} encoders in each arm measure the desired angle.

When instructed to point to a target \aztext{} and \eltext{}, the encoders ensure that the \aztext{} and \eltext{} axes are set to the instructed angles.
However, in general, the  beam does not point toward the same instructed \aztext{} and \eltext{} on the sky because imperfections distort the telescope structure.
These distortions can be caused by physical effects including: gravitational flexure, tilts of the structural axes, collimation errors, and encoder offsets.
The distortions are modeled to quantify adjustments to the angles instructed to the \aztext{} and \eltext{} encoders.
These adjustments comprise the ``pointing model.''

In this section, we discuss the pointing model employed on the SPT, how the model changes and causes pointing error, how we deal with those errors for normal SPT operation, and how these errors cause issues for EHT observations.

\subsection{The Pointing Model}
\label{sec:pointing_model}

The pointing model used on the SPT is based on a pointing model developed for the Cosmic Background Imager \cite{padin02}.
Although that pointing model is not published in a scientific journal, it is accessible publicly online in an internal memo written by Martin Shepherd.\footnote{\url{https://sites.astro.caltech.edu/~mcs/CBI/pointing/index.html}}
This section provides a complete description of the model as previously implemented in the SPT control software.

Locations of targets on the sky are typically defined first in astrometric coordinates, such as equatorial right ascension and declination.
The beginning step of a pointing model is converting from astrometric coordinates to topographic \aztext{} and \eltext{} assuming the telescope is a perfect instrument with no physical deformation.
This conversion includes accounting for the location of the telescope on Earth and for atmospheric effects such as refraction and aberration.
Refraction corrections are implemented in the SPT control software using the \textsc{SLALIB} library \cite{wallace94}.
In this work, we consider this coordinate transformation a solved problem, and we focus on the mechanical deformations resulting from an imperfect instrument.

The pointing model is a set of equations applied in succession to correct for a set of modeled imperfections.
We denote the target \aztext{} and \eltext{} instructed to the encoders after the $i$-th step of correction as $\az_i$ and $\el_i$, respectively.
Each correction is calculated using the previously corrected \aztext{} $\az_{i-1}$ and \eltext{} $\el_{i-1}$.
The full pointing model consists of $N$ corrections, resulting in \aztext{} $\az_N$  and \eltext{} $\el_N$ instructed to the encoders such that the telescope points to the target \aztext{} $\az_0$ and \eltext{} $\el_0$.

Once we have a target \aztext{} and \eltext{}, the first step in the pointing model is correcting for gravitational flexure.
The SPT \eltext{} axis supports two large, massive structures whose supports are at risk of sagging due to gravity: the primary mirror and the receiver cabin.
These structures are roughly orthogonal to each other.
They are aligned such that when the telescope is pointed at the horizon, the center of mass of the primary mirror support structure is aligned roughly vertically with the \eltext{} axis and flexure is negligible, while the center of mass of the receiver cabin support structure is aligned roughly horizontally and flexure is maximal.
When the telescope is pointed at zenith, the center of mass of the primary mirror support structure is aligned roughly horizontally with the \eltext{} axis and flexure is maximal, while the center of mass of the receiver cabin support structure is aligned roughly vertically and flexure is negligible.

Thus we model elevation-dependent flexure corrections with two terms: one that varies like $\sin{\el}$ and one that varies like $\cos{\el}$.
The correction applied is
\begin{equation}
\begin{gathered}
	\el_1 = \el_0 - \azero \sin{\el_0} - \aone \cos{\el_0} \\
	\az_1 = \az_0
\end{gathered}
\label{eq:flex}
\end{equation}
where the parameter $\azero$ specifies the amount of primary mirror flexure and the parameter $\aone$ specifies the amount of receiver cabin flexure.
Typical operation assumes values of $\azero = 29\farcs03$ and $\aone = -50\farcs91$.
Unless otherwise noted, in this section we report typical operation values which were determined from a dedicated pointing measurement excercise described later in Section~\ref{sec:spt_point}.

The second step in the pointing model is correcting for any tilt in the vertical \aztext{} axis of the telescope mount.
This correction accounts for the fact that the \aztext{} axis is not perfectly parallel to zenith on the spheroidal model of Earth with which we calculate our target \aztext{} and \eltext{}.
We posit two potential physical reasons: (1) the \aztext{} axis is aligned as closely as possible to local gravity, but local gravity varies from the spheroidal model of Earth, and (2) the alignment of the \aztext{} axis is distorted by imperfections in the \aztext{} bearing, a varying center of mass, and a support system which gradually sinks into the ice foundation.

We model tilting of the \aztext{} axis with two more parameters: $\atwo$ and $\athree$.
Suppose the \aztext{} axis is tilted away from zenith by a magnitude $\rho$ toward the \aztext{} direction $\Omega$.
Then, we define $\athree$ as the magnitude of tilt in the direction opposite of $\Omega$, such that a positive $\athree$ corresponds to a higher \eltext{} than expected.
We can define $\atwo$ as the magnitude of the tilt in a direction perpendicular to the \aztext{} axis vector and the positive $\atwo$ direction such that the directions of positive $\athree$, $\atwo$, and $\az$ satisfy the right hand rule in that order.
It can then be shown that
\begin{equation}
\begin{gathered}
	\Omega = - 180^\circ - \ATAN{
		\frac{\SIN{\atwo \cos{\athree}}}
		{\COS{\atwo \cos{\athree}} \sin{\athree}}
	}\\
	\rho = \ACOS{ \cos{\athree} \COS{\atwo \cos{\athree}} }.
\end{gathered}
\end{equation}
With this notation, we can describe the corrections applied to \eltext{} (\elmath) and \aztext{} (\azmath) as:
\begin{equation}
\begin{gathered}
	\el_2 = \ASIN{ \sin{\el_1} \cos{\rho} - \cos{\el_1} \sin{\rho} \sin{\omega} } \\
	\az_2 = \Omega - \ATAN{ \frac{\cos{\omega} \cos{\el_2}}{- \cos{\rho} \sin{\omega} \cos{\el_2} - \sin{\rho} \sin{\el_2}} } \\
\end{gathered}
\label{eq:aztilt}
\end{equation}
where $\omega \equiv \Omega - 90^\circ - \az_1$.
The true values of $\atwo$ and $\athree$ change over time as the ice foundation settles.
At the start of every day of observations, we directly measure $\atwo$ and $\athree$ by rotating the telescope fully in \aztext{} and monitoring readings from a biaxial tiltmeter \cite{carlstrom11}.
On 2024 January 01, the values were measured to be $\atwo = 3\arcmin 36\farcs60$ and $\athree = -4\arcmin 03\farcs89$.

The third step in the pointing model is correcting for any tilt in the horizontal \eltext{} axis of the telescope mount.
This correction accounts for the fact that the \eltext{} axis may not lie perfectly oriented within the \aztext{} plane, which can occur in two ways.
First, the \eltext{} axis may be rotated about the \aztext{} axis.
This rotation is indistinguishable from an \aztext{} encoder offset, which we correct later.
Alternatively, the \eltext{} axis may be rotated out of the \aztext{} plane.

This rotation can be modeled with one parameter we define as $\afour$, which measures the magnitude of the \eltext{} axis tilting out of the \aztext{} plane.
It can be shown that the corrections applied to \eltext{} (\elmath) and \aztext{} (\azmath) are equal to
\begin{equation}
\begin{gathered}
	\el_3 = \ASIN{ \frac{\sin{\el_2}}{\cos{\afour}} } \\
	\az_3 = \az_2 - \ASIN{\tan{\el_3} \tan{\afour}}.
\end{gathered}
\label{eq:eltilt}
\end{equation}
Typical operation assumes a value of $\afour = -28\farcs18$.
However, we expect this parameter to vary on relatively short timescales, as we discuss later in Section~\ref{sec:var_point}.

The fourth step in the pointing model is correcting for errors in collimation.
This correction accounts for the fact that the focal plane may not be perfectly perpendicular to the direction of light hitting it from the tertiary mirror of the telescope.

This correction can be modeled with two parameters we define as $\asix$, which corresponds to an offset in the \eltext{} direction, and $\afive$, which corresponds to an offset in the \xeltext{} direction.
We can convert these offsets to polar coordinates on the celestial sphere with magnitude $\mu$ and direction $\theta$, with $\theta = 0$ pointing towards the direction of increasing \eltext{} and $\phi$ increasing anti-clockwise.
We calculate those values as follows:
\begin{equation}
\begin{gathered}
	\mu = \ACOS{ \cos{\afive} \cos{\asix} } \\
	\theta = \ATAN{ \frac{\sin{\afive}}{\tan{\asix}} }
\end{gathered}
\end{equation}
This notation simplifies the equations describing the corrections applied to \eltext{} (\elmath) and \aztext{} (\azmath):
\begin{equation}
\begin{gathered}
	\el_4 = \ASIN{ \cos{\mu} \sin{\el_3} + \sin{\mu} \cos{\el_3} \cos{\theta} } \\
	\az_4 = \az_3 + \ATAN{ \frac{\sin{\mu}\sin{\theta}}{\cos{\el_3}\cos{\mu} - \sin{\el_3}\sin{\mu}\cos{\theta}} }.
\end{gathered}
\label{eq:coll}
\end{equation}
Typical operation assumes a value of $\afive = 4\arcmin 11\farcs17$ and $\asix = -14\arcmin 13\farcs36$.
However, we expect $\asix$ to vary on relatively short timescales, as we discuss later in Section~\ref{sec:var_point}.

The final step in the pointing model is correcting for errors in encoder offset.
That is, the directions towards which the encoders read $\az = 0\degree$ and $\el = 0\degree$ may not agree with the true topographic origin.
Furthermore, an offset in the \aztext{} encoder accounts for any constant timing offset in the control software when monitoring local sidereal time because of the location at the geographic South Pole.

This correction can be modeled as two parameters $\aseven$ and $\azzero$, which correspond to the offsets in the origins of the \eltext{} encoder and the \aztext{} encoder, respectively.
These corrections are direct additions to instructed encoder \eltext{} (\elmath) and \aztext{} (\azmath); i.e.
\begin{equation}
\begin{gathered}
	\el_5 = \el_4 + \aseven \\
	\az_5 = \az_4 + \azzero.
\end{gathered}
\label{eq:off}
\end{equation}
For small values of $\asix$, Eq.~(\ref{eq:coll}) simplifies to $\el_i - \el_{i-1} \approx \asix$.
Therefore, $\aseven$ is degenerate with $\asix$ to first order, so typical operation assumes a value of $\aseven = 0\arcsec$.
However, $\azzero$ is not degenerate with $\afive$, and typical operation assumes a value of $\azzero = -20\arcmin 15\farcs46$.

\begin{figure*}[]
	\centering
	\includegraphics[width=0.9\linewidth]{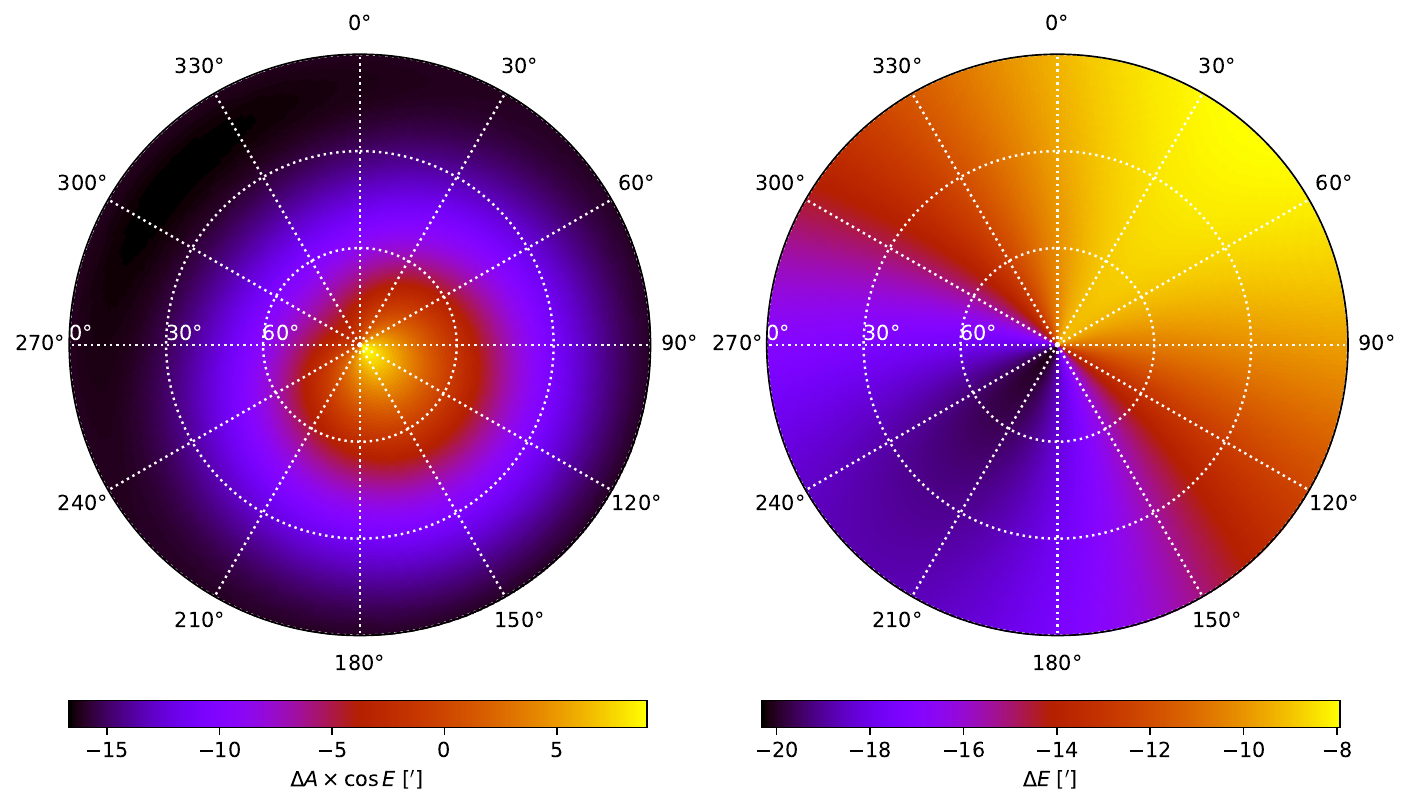}
	\caption{
		\small
		The net impact of typical corrections \textit{(left)} $\Daz \equiv \az_5 - \az_0$ to \aztext{} and \textit{(right)} $\Del \equiv \el_5 - \el_0$ to \eltext{} applied via the pointing model, which can be calculated with Eqs.~(\ref{eq:flex}), (\ref{eq:aztilt}), (\ref{eq:eltilt}), (\ref{eq:coll}), and (\ref{eq:off}).
		Positions $\az_5$ and $\el_5$ are instructed to the \aztext{} encoder and \eltext{} encoder, respectively, to account for imperfections in the telescope structure so that the telescope points to target positions $\az_0$ and $\el_0$.
		Pointing model corrections are functions of \aztext{} (the angular axis on the plots), \eltext{} (the radial axis on the plots), and pointing model parameters.
		The plotted corrections use values of pointing model parameters assumed during a typical observation around 2024 January 01.
		The plotted $\Daz$ corrections are multiplied by $\cos{\el}$ to avoid divergence at high \eltext{}s and resemble approximate corrections to \xeltext{}.
	}
	\label{fig:pointing_model}
\end{figure*}

The pointing model is composed of the five corrections modeled by Eqs.~(\ref{eq:flex}), (\ref{eq:aztilt}), (\ref{eq:eltilt}), (\ref{eq:coll}), and (\ref{eq:off}).
The net impact of these corrections, using the values assumed during typical observations, is shown in Figure~\ref{fig:pointing_model}.
The corrections can be simplified using small angle approximations for small pointing model parameter values, resulting in commutative corrections that are accurate for most values of \aztext{} and \eltext{}.
However, the equations described here reflect the full pointing model as previously implemented on the telescope control system for general use. 

\subsection{Time-Varying Pointing Model Parameters}
\label{sec:var_point}

We expect the optimal values of some of the pointing model parameters defined in Section~\ref{sec:pointing_model} to change over time.
For instance, the foundation settles into the ice and causes $\atwo$ and $\athree$ to change gradually over years.
We are able to directly monitor, and thus correct for, those changing parameter values.

However, not all changing parameters can be monitored directly.
The primary effect we expect to be unable to monitor is thermal deformation of the telescope structure.
In particular, there is a temperature gradient between the warm base of the rotating roof (which covers the moderately heated control room) and the cold primary dish and receiver cabin (the structural components of which are completely exposed to the elements).
The two yoke arms that support the \eltext{} axis extend through the rotating roof; these arms are expected to exhibit the strongest temperature gradient.
Although the yoke arms are covered by a foam insulation to mitigate weather effects, we expect the temperature gradients along the yoke arms to change as the telescope changes its position and as weather conditions change.
Subsequently, we expect the pointing accuracy of the telescope to change.

Some radio telescopes directly monitor changing pointing model parameters by using small optical star pointing telescopes.
Optical guide telescopes were implemented on SPT as described in \citet{carlstrom11}. 
The utility of the optically determined pointing for radio pointing was limited, presumably due to variable collimation offsets between the optical boresites and mm beam caused by local deformations of the telescope structure.
Given the ability to correct pointing in CMB observations after the fact, the optical system was not used, and the optical guide telescopes were removed when their mounting interfered with the installation of additional shielding structures on the SPT.

\begin{figure*}[]
	\centering
	\includegraphics[width=1\linewidth]{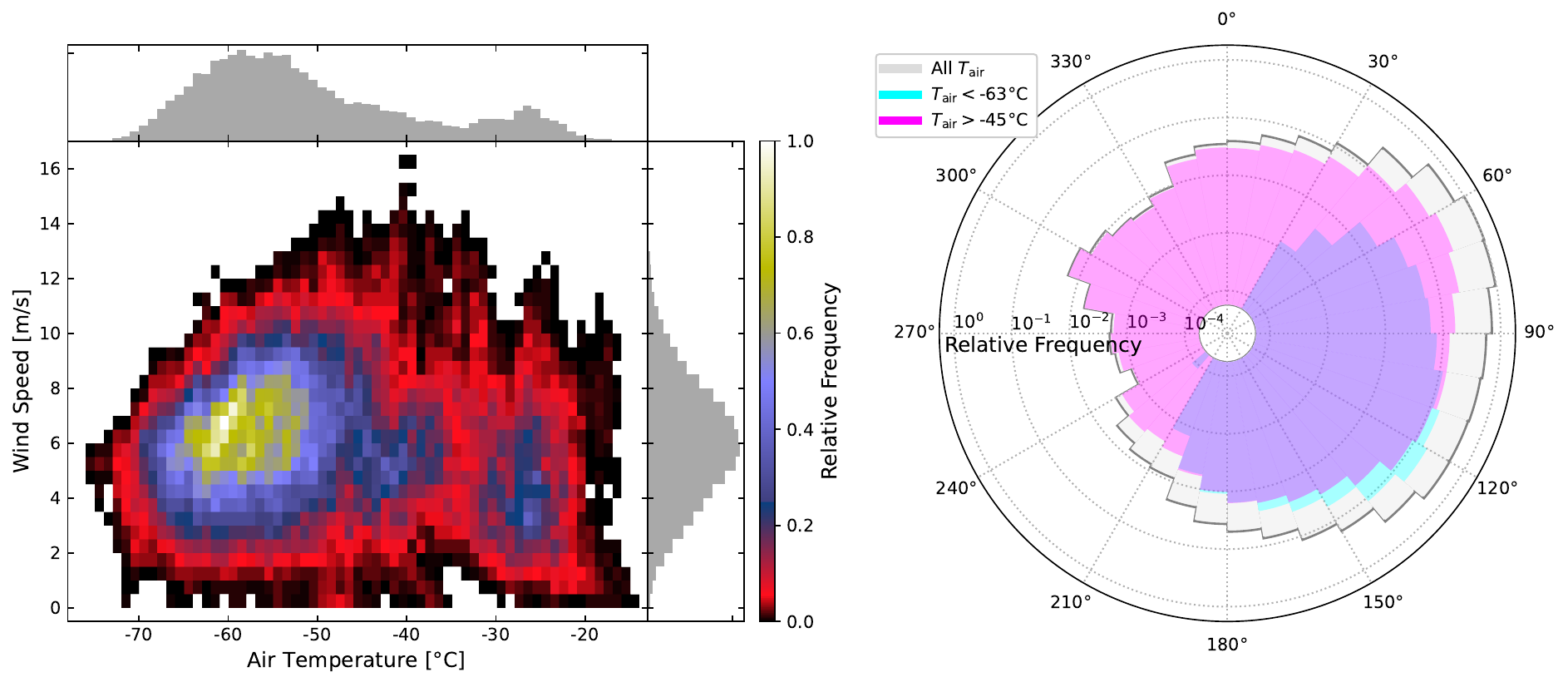}
	\caption{
		\small
		Historical weather conditions at the SPT showing \textit{(left)} a two-dimensional histogram of  wind speed versus ambient air temperature and \textit{(right)} a histogram of the \aztext{} direction from which the wind blows.
		The plotted data were measured by a weather station on a nearby roof between 2019 February and 2023 September, and they are the same data described in Section~\ref{sec:dataset}.
	}
	\label{fig:weather}
\end{figure*}

Weather conditions at the South Pole are extreme but generally stable.
Figure~\ref{fig:weather} shows histograms of typical weather conditions as measured by an SPT-operated weather station on a nearby roof.
Note that these measurements generally agree with those reported in \citet{lazzara12}.
The air temperature is generally around -$65\degree$C to -$50\degree$C, especially during the cold austral winter, although it gets warmer during the austral summer.
Wind blows on average around 6~m/s, though it can gust upwards of 12~m/s.
The wind blows predominantly from one direction, around $\az \sim 65\degree$, although this shifts during the coldest weather to $\az \sim 125\degree$.

We can model thermal deformation of the yoke arms via two dominant modes.
The first mode occurs when one yoke arm is shorter or longer than the other.
Since the yoke arms support the \eltext{} axis, differential changes in the length of the arms cause a rotation of the \eltext{} axis out of the \aztext{} plane, which is exactly the imperfection described by the $\afour$ parameter.
When the variation of $\afour$ is small, we can use small angle approximations on Eq.~(\ref{eq:eltilt}) to show that
\begin{equation}
	\Delta\afour \approx - \Daz / \tan{\el}.
\label{eq:aza4}
\end{equation}
That is, a time-varying imperfection of the kind described by $\afour$ causes pointing errors in \aztext{}.
We expect that the yoke arms are able to expand or contract by a few millimeters.
Given the geometry of the structure, one arm deforming just 0.5~mm more than the other (corresponding with a temperature difference of $\sim 12\degree$C) induces a deformation that can be accounted for by a $30\arcsec$ change in $\afour$.

The second mode occurs when either the front of the yoke arms expands or contracts more than the back.
This deformation causes errors in pointing that can be described mathematically equivalently to the definition of $\asix$.
When the variation of $\asix$ is small, we can use small angle approximations on Eq.~(\ref{eq:coll}) to show that
\begin{equation}
	\Delta\asix \approx \Del.
\label{eq:ela6}
\end{equation}
I.e. a time-varying imperfection of the kind described by $\asix$ causes pointing errors in \eltext{}.
Given the geometry of the structure, the front of the arms deforming just 0.25~mm more than the back (corresponding with a temperature difference of $\sim 6\degree$C) induces a deformation that can be accounted for by a $31\arcsec$ change in $\asix$.

\begin{figure*}[!htp]
	\centering
	\subfloat{\includegraphics[width = 0.82\linewidth]{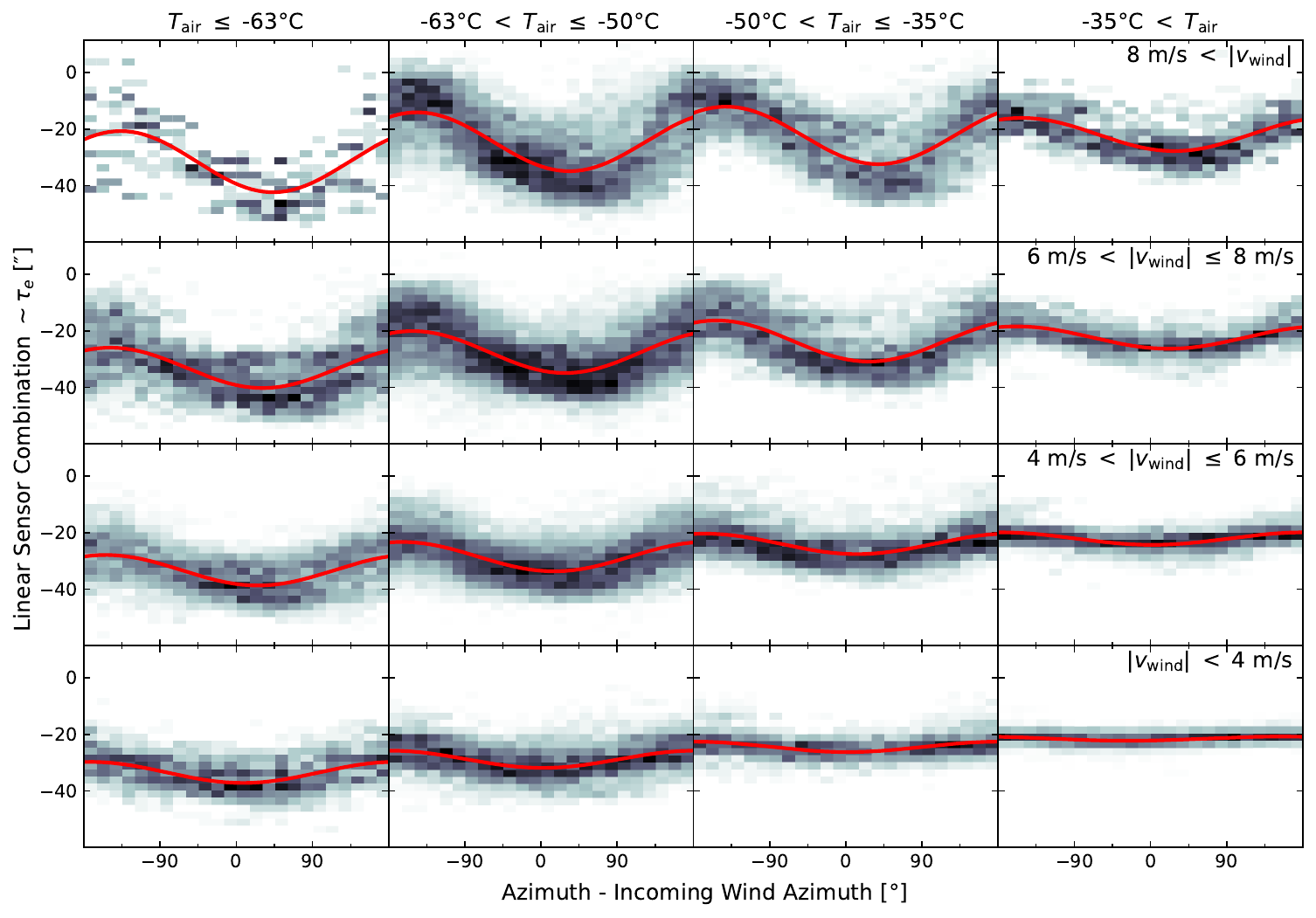}}\\
	\subfloat{\includegraphics[width = 0.82\linewidth]{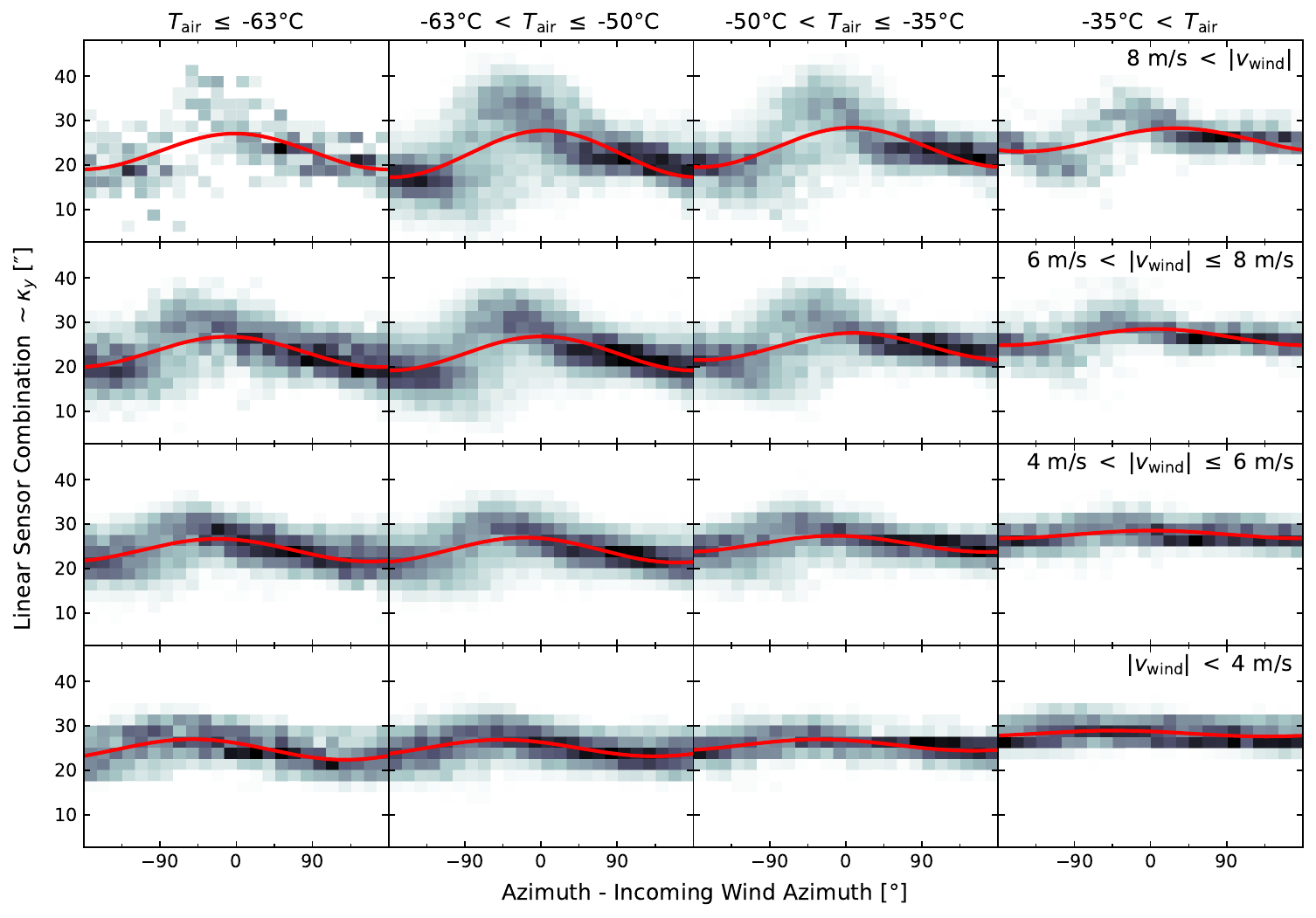}}
	\caption{
		\small
		Effect of weather and observing conditions on deformations in the telescope yoke arms.
		The data plotted are the same dataset described in Section~\ref{sec:dataset}.
		The data are binned into different ranges of air temperature and wind speed to show the effect of weather conditions, then they are plotted as two-dimensional histograms.
		The yoke arm deformations are measured by linear sensors described in Section~\ref{sec:features}; we plot linear combinations of the sensors that we expect correspond with varying \textit{(top)} $\afour$ and \textit{(bottom)} $\asix$.
		The deformations depend on the direction the telescope points relative to incoming wind, and we overplot a best-fit sinusoidal function in red.
	}
	\label{fig:linsens_weather}
\end{figure*}

\begin{figure*}[!ht]
	\centering
	\includegraphics[width=0.8\linewidth]{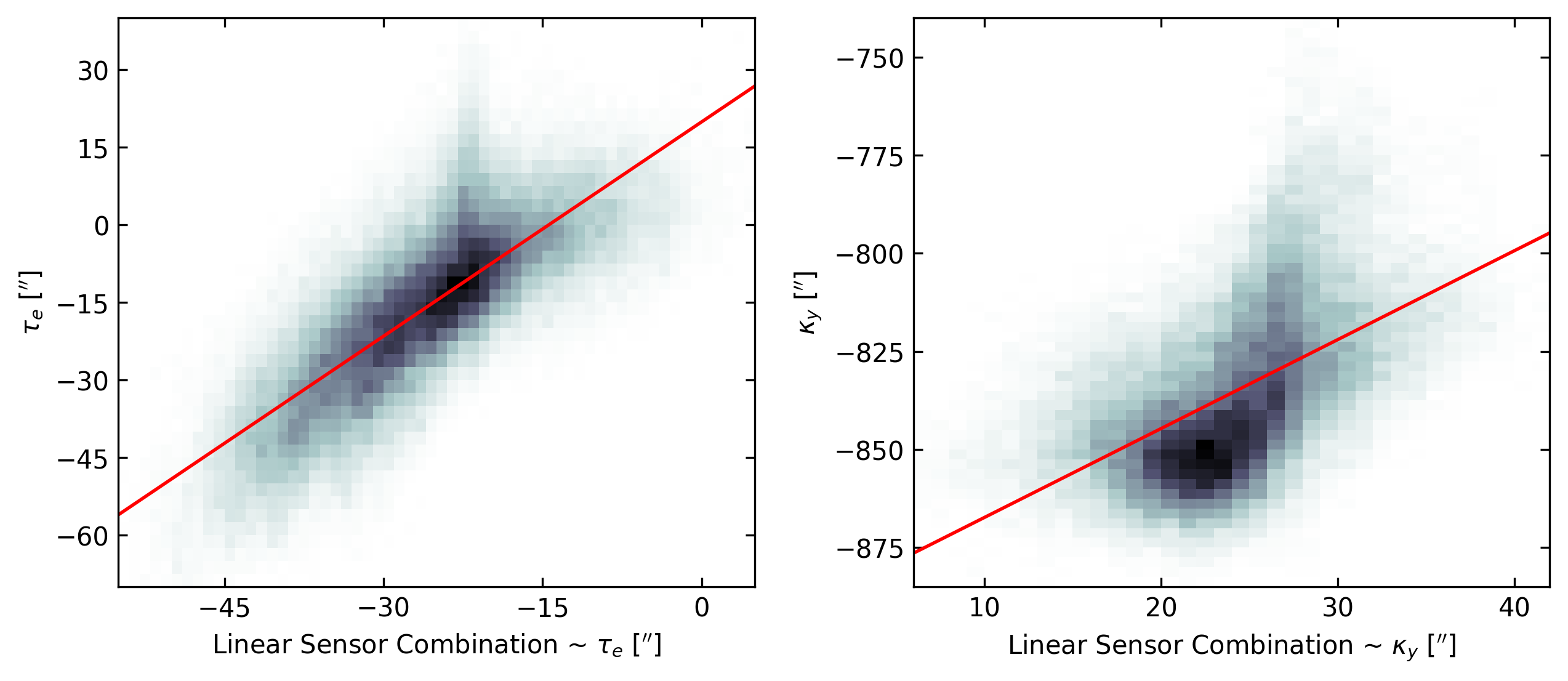}
	\caption{
		\small
		Two-dimensional histogram showing correlation between optimal pointing model parameter values and yoke arm deformations.
		The yoke arm deformations are measured by linear sensors described in Section~\ref{sec:features}; we plot linear combinations of the sensors that we expect correspond with varying \textit{(left)} $\afour$ and \textit{(right)} $\asix$.
		The data plotted are the same dataset described in Section~\ref{sec:dataset}.
		We overplot a best-fit linear function in red.
	}
	\label{fig:a4a6_vs_linsens}
\end{figure*}

As weather conditions change and as the SPT points to different locations, the structure deforms in a way described by varying $\afour$ and $\asix$.
Historical pointing corrections suggest that $\afour$ and $\asix$ vary on short (minute to hourly) and long (seasonal) timescales, but we do not notice other long-term trends.
We cannot directly monitor variations of $\afour$ and $\asix$, but we do have sensors, described later in Section~\ref{sec:features}, that measure length changes in the yoke arms.
We can take linear combinations of these sensors that geometrically reflect deformations of the kinds parametrized with $\afour$ and $\asix$; these calculations can serve as a proxy for direct monitoring.
Figure~\ref{fig:linsens_weather} shows how these calculations typically vary due to different observing and weather conditions.
In general, these deformations depend on the wind direction relative to the direction the telescope is pointed, and the deformations are larger with higher wind speeds and lower air temperature.
The varying deformations introduce time-dependent pointing errors in \aztext{} and \eltext{}.
Figure~\ref{fig:a4a6_vs_linsens} shows how $\afour$ and $\asix$ correspond to measured changes in the sensors.
Monitoring these sensors can correct much of, but not all of, the variation in observed pointing errors: simple linear regression achieves $R^2=0.57$ for $\afour$ and only $R^2=0.28$ for $\asix$.
We require more complex modeling such as ML to reach our goal accuracy.

\subsection{Pointing for Typical SPT Observations}
\label{sec:spt_point}

The observations made with the SPT predominantly target making low-noise measurements of the CMB, which is accomplished by repeatedly observing the same region of the sky.
Each target region of the sky is on the order of hundreds of square degrees in size, and the SPT-3G focal plane provides an instantaneous field of view of about two square degrees.
Observations are performed by scanning back and forth in \aztext{}, stepping in \eltext{}, then repeating until the entire target sky region is observed.
Each entire observation takes approximately two hours to complete.

Typical observations are made using a set of mostly-static ``online'' pointing model parameters.
The tilt parameters $\atwo$ and $\athree$ are measured at the start of each observation day and assumed constant throughout the day.
The other parameters assume historical values that are accurate up to effects of observing mode and time variation.
These values were measured with dedicated observations of known astronomical sources.
The SPT was used to scan over many sources in succession, and observed pointing errors were measured by referencing the expected locations of the sources.
This dataset was then used to fit the pointing model equations for parameter values that minimized pointing error.

Typical observations are analyzed by making maps of the sky region.
The observational data consist of time-ordered data (TOD) stored for each of the thousands of detectors on the focal plane.
To make maps, the location to which each detector was pointed at each time sample is calculated using the pointing model.
The TOD are binned into pixels on the sky based on the pointing calculated for each detector.
All time samples from all detectors pointed at a given pixel are averaged to produce a single value for that pixel.
The collection of all these sky pixels comprise the resulting map.
One could change the pointing model parameters used to calculate pointing during the mapmaking process.
Changing the pointing model parameters in this way would cause the calculated positions of the detectors to change and thus shift and stretch the resulting map around the sky.

One can estimate the typical uncorrected pointing error by using the online pointing model parameters to make maps.
However, these maps are typically not created because we expect relatively poor pointing accuracy with the online pointing model parameters.
That is, because we expect $\afour$ and $\asix$ to change in time as described in Section~\ref{sec:var_point}, we expect to see known bright sources appearing at locations different from their known positions.
We describe the magnitude of pointing offsets by calculating the root mean square error (RMSE) of observed positions of bright sources from their expected positions, and we use small-angle approximations to estimate an offset in cross-elevation as equivalent to an offset in \aztext{} times the cosine of \eltext{}.
If we made maps using the online pointing model parameters, we expect the median of all single-observation RMSE values to be about 12\farcs9 in \xeltext{} and 19\farcs5 in \eltext{}, or about 27\farcs1 in distance along a great circle.
These are the uncorrected pointing errors of typical SPT observations without accounting for time-varying $\afour$ and $\asix$.

Because we can remake maps using different pointing model parameters, we can correct after-the-fact for observed pointing errors.
We call these corrections ``offline'' pointing corrections.
The first step in offline pointing corrections utilizes measurements of calibration sources in between each observation.
The calibration sources are typically HII regions, or bright galactic star-forming regions.
During these measurements, the SPT rasters over a calibration source, and we process the TOD to create a map of the source using the online pointing model parameters.
We calculate an observed position of the calibration source by computing its flux-weighted centroid, and we compare the observed position to the expected position.
We find the values of $\afour$ and $\asix$ that minimize the measured pointing offset.
Because the time of the calibration source measurement is close to the time of the observation, we expect these values of $\afour$ and $\asix$ to be close to the optimal values for the observation.
After making maps with this first-order offline correction, the median of all single-observation RMSE values for bright sources is 10\farcs3 in \xeltext{} and 10\farcs7 in \eltext{}, or about 16\farcs8 in distance along a great circle.

The second step in offline pointing corrections involves making maps of the actual observation and correcting the positions of bright sources appearing in the map.
The map is made using the values fit for $\afour$ and $\asix$ from the calibration source measurements.
We select known bright sources in the map that appear with measured S/N $>10$, of which there are typically 3-20 per observation.
We do this only using maps made with the SPT-3G 95~GHz band, in which we measure the sources with the highest S/N.
For each selected source, we fit a two-dimensional Gaussian and calculate an offset from its expected position.
All sources are used to find the values of $\afour$ and $\asix$ that minimize the observed offset of the sources on average.
This is the final step of offline pointing corrections, after which final maps are made with the updated parameters.
After making maps with this second-order offline correction, the median of all single-observation RMSE values for bright sources is 2\farcs7 in \xeltext{} and 4\farcs0 in \eltext{}, or about 5\farcs1 in distance along a great circle.
These are the final pointing errors in typical SPT CMB observations.

\begin{figure}[!ht]
	\centering
	\includegraphics[width=0.92\linewidth]{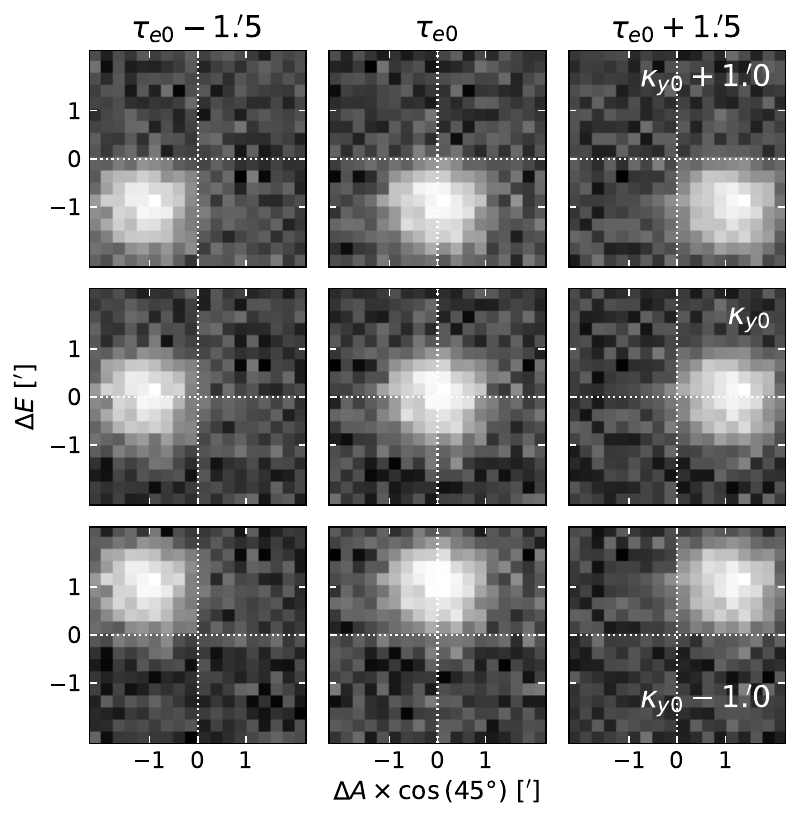}
	\caption{
		\small
		The measured position of a source changes when the values of $\afour$ and $\asix$ are changed during mapmaking.
		The ideal values, $\afour_0$ and $\asix_0$, respectively, minimize the offset of the source from the expected position \textit{(center)}.
		This source appears at $\el = 45\degree$, so changing $\afour$ by $\pm 1\farcm5$ would move the telescope by $\Daz \times \cos{\el} \approx \mp 1\arcmin$ according to Eq.~(\ref{eq:aza4}), which causes the source to appear to move by $\Daz \times \cos{\el} \approx \pm 1\arcmin$.
		Changing $\asix$ by $\pm 1\arcmin$ would move the telescope by $\Del \approx \pm 1\arcmin$ according to Eq.~(\ref{eq:ela6}), which causes the source to appear to move by $\Del \approx \mp 1\arcmin$.
		At the geographic South Pole, change in R.A.~$\Delta\alpha \approx \Daz$ and change in decl.~$\Delta\delta \approx -\Del$, so these changes cause the source to appear to move by $\Delta\alpha \times \cos{\delta} \approx \pm 1\arcmin$ and $\Delta\delta \approx \pm 1\arcmin$, respectively.
	}
	\label{fig:thumbnail_a4a6}
\end{figure}

Figure~\ref{fig:thumbnail_a4a6} shows an example of changing the offline pointing model parameter values during mapmaking for one source.
As we vary the values of $\afour$ and $\asix$ from the optimal values determined during the fitting procedure, the measured position of the source moves away from its expected position.

Even with perfect pointing, we would still measure some RMSE in observed bright source positions.
This is because of statistical uncertainty while measuring the position of a source.
Suppose the shape of the beam can be modeled by a two-dimensional Gaussian with some full width half max (FWHM).
Then, the uncertainty in the measured position along each axis is approximately 0.6 times the beam FWHM divided by the source S/N (see, for instance, Appendix B of \citealt{ivison07} for a derivation).
The FWHM of the SPT-3G camera is about 1\farcm57 at 95~GHz \cite{sobrin22}.
For the choice of selecting sources with S/N $> 10$, we expect an uncertainty of $\lesssim 5\farcs7$ in measurements of offsets along each axis.

The final pointing errors in typical SPT observations are well below the size of the beam.
Thus, the offline pointing corrections described here are sufficient for precision cosmology and other scientific analysis of SPT data.

\subsection{Pointing for EHT Observations}
\label{sec:eht_point}

EHT observations with the SPT require stricter pointing than normal SPT CMB survey data.
The EHT receiver installed in the SPT is separate from the SPT-3G receiver and consists of a single-pixel superconducting mixer receiver \cite{kim18b}.
Since there is only a single receiver pixel and scientific observations do not include raster scanning, the SPT must point with accurate online parameters to measure the target signal.
Pointing corrections cannot be implemented offline as they are for CMB observations.

Additionally, pointing the SPT for EHT observations is difficult because we expect there to be different optimal values for pointing model parameters than during CMB observations and we must determine those values.
The EHT and SPT-3G receivers are housed simultaneously in the receiver cabin, although they cannot observe in tandem as they use separate, conflicting optical mirrors.
Thus, we expect constant offsets in $\afive$ and $\asix$ to be the primary difference between EHT and SPT-3G operating modes. 
The mirrors used for EHT are installed for each observation campaign and removed thereafter, so we expect the optimal values of $\afive$ and $\asix$ to change between each installation.
In addition, we still expect $\afour$ and $\asix$ to vary due to structural deformations as described in Section~\ref{sec:var_point}.

To aid in collecting pointing offset data, the EHT receiver has a digital spectrometer that we use to make observations of either sources with prominent carbon monoxide emission lines or bright thermal sources like planets.
The spectrometer is used to raster scan over these sources and create maps which are then fit with a two-dimensional Gaussian to measure pointing offsets.

Historically, we have approached solving the difficulties with pointing by performing a set of observations over known calibration sources, during which we repeatedly cycle through a list of sources and scan each source.
EHT observing campaigns typically consist of about one week of observing sessions, scattered over roughly two weeks to select for the best aggregate observing conditions at all participating sites.
In the few days of setup prior to the first possible observing session, we perform pointing observations with the SPT, typically for at least 24 hours.
As the sources move around the sky throughout this time, we sample many \aztext{} positions for each source and use a variety of sources to sample different \eltext{}s.
We use these observations to fit for constant values of $\afour$, $\afive$, and $\asix$ that are accurate up to changing conditions as described in Section~\ref{sec:var_point}.

One session of an EHT observing campaign can last many hours and occur many days after the pointing observations.
To account for varying pointing model parameters, we also perform source observations scattered throughout the session.
The session is usually very busy with frequently scheduled measurements of scientific targets.
When time allows, we point the telescope to a pointing source near an upcoming science target, perform a raster scan pointing observation, fit for an observed offset, and apply the offset as an additional correction to the online pointing model.
This strategy suffices when there is a pointing source near the science target and when we have time to scan over a pointing source.
These conditions are not always met, in which case pointing accuracy suffers.

Historically, the RMSE of residual pointing offsets measured during pointing observations after fitting for pointing model parameters is about 10\arcsec{} in \xeltext{} and 10\arcsec{} in \eltext{}, or about 14\arcsec{} in distance along a great circle.
The magnitudes of these errors are comparable to those remaining after first-order offline pointing corrections for typical SPT observations.
This similarity is expected; after determining optimal pointing model parameter values with either a calibration source during CMB observations or with pointing sources during EHT observations, we expect the telescope to deform in the same way to changing weather conditions.

For SPT observations to meaningfully contribute to the EHT array, we require the online pointing accuracy to be some small fraction of the beam size.
The EHT receiver is dual-frequency, facilitating observations at 230~GHz and 345~GHz.
Currently, only the 230~GHz receiver is operational, as the 345~GHz receiver for the SPT is still under development.
We estimate the resolution of the 230~GHz receiver by fitting a two-dimensional Gaussian to observations of Saturn with high S/N, resulting in a best-fit Gaussian with $\sim 45\arcsec$ FWHM.
We note that this is roughly consistent with a 6-m effective aperture diameter as reported in \citet{eht19b}.\footnote{
	Assuming a 6-m diameter mirror with top-hat illumination, the Rayleigh criterion describes the beam as an Airy disk with first minima at $1.22 \frac{\lambda}{d} = 55\arcsec$, where $\lambda$ is the observing wavelength and $d$ is the effective diameter.
	A Gaussian function that most closely approximates this shape has a $\sim 45\arcsec$ FWHM.
}
The 345~GHz receiver will have a smaller beam than the 230~GHz receiver according to the Rayleigh criterion and thus stricter pointing requirements.
We model the beam of the 345~GHz receiver as a two-dimensional Gaussian with 30\arcsec{} FWHM.

Our goal is to enable pointing accuracy that is sufficient to facilitate the use of the 345~GHz receiver.
Assuming a beam profile, we can express any pointing error in terms of a loss in detector response.
Given our model of the 345~GHz receiver, a source offset along a great circle by 5\farcs9 would have a detector response of 90\%.
If errors in \eltext{} and in \xeltext{} are Gaussian-distributed, then errors along a great circle are Rayleigh-distributed.
A Rayleigh distribution with 5\farcs9 median error corresponds with 5\arcsec{} RMSE in \eltext{} and in \xeltext{}.
This is our target goal for this work.

\section{Machine Learning Training}
\label{sec:ml_training}

To achieve the pointing accuracy requirements described in Section~\ref{sec:eht_point}, we need to estimate in real time the time-varying pointing model parameter values which cannot be measured directly, as described in Section~\ref{sec:var_point}.
We aim to do this by training an ML model that can use information about the current telescope and weather conditions to estimate those parameters.
This section describes the data, architecture, and methods used during model training.

\subsection{Target Variables}

We aim to correct for \aztext{} and \eltext{} pointing offsets in real time.
We expect these offsets are caused primarily by time-varying deformations that can be described by the pointing model parameters $\afour$ and $\asix$, which functionally affect \aztext{} and \eltext{} offsets via Eqs.~(\ref{eq:aza4}) and (\ref{eq:ela6}), respectively.
If we can accurately estimate the true values of $\afour$ and $\asix$, then we can update those values in the corrections used online by the telescope control software.
Even if the offsets are not caused solely by deformations that are described perfectly by $\afour$ and $\asix$, we can correct all \aztext{} and \eltext{} offsets with these parameters, so predicting just these parameters is a sufficient solution.
This is a supervised regression problem in which $\afour$ and $\asix$ are our target variables.

\subsection{Feature Variables}
\label{sec:features}

The most obvious included input features are the target \aztext{} and \eltext{}, for three main reasons.
First, we know that the pointing model depends on target \aztext{} and \eltext{}.
If we have not perfectly measured the static pointing model parameters, we would introduce \aztext{} and \eltext{} dependencies into the measured source offsets.
Second, we aim to estimate the $\afour$ parameter, which we know affects \aztext{} offsets depending on target \eltext{}, as described by Eq.~(\ref{eq:aza4}).
Third, structural deformations of the same functional form as $\afour$ and $\asix$ may not be the only time-varying deformations which cause pointing errors.
Since we aim to correct all \aztext{} and \eltext{} offsets by predicting just these two parameters, we include \aztext{} and \eltext{} as features to pick up functionally different deformations.

Some sensors are part of the telescope structure and allow us to monitor the state of the telescope.
One such type of sensor is a ``linear displacement sensor'' mounted inside the yoke arms \cite{carlstrom11}.
In total, four linear sensors, one in the front and back of each yoke arm, measure changes in the height of the yoke arms relative to a stiff, carbon-fiber-reinforced-plastic (CFRP) reference frame.
The linear sensors provide a direct measurement of structural deformations that we expect to be primary contributors to variations in $\afour$ and $\asix$. 

Another type of sensor on the telescope structure is thermometry.
A total of 60 thermometers are scattered throughout the structure, specifically on the pedestal, inside the yoke arms, inside the receiver cabin, and along the structural boom that extends from the primary mirror to the receiver cabin.
The pedestal is the support in the ice at the base of the telescope; we do not expect this structure to change meaningfully and do not include these thermometers as features.
The inside of the receiver cabin is kept at nearly constant temperature with heaters, and these thermometers sit on non-structural components; we do not include these thermometers as features.
We use the 33 remaining functional thermometers as features.

A weather station monitors weather conditions from the roof of a laboratory building located about 20~m away.
Of particular relevance here, the station records the ambient temperature of the outside air, the azimuthal direction of the oncoming wind, and the speed of the wind.
These data are included as features.

We make a modest attempt at feature engineering prior to model training.
For instance, we expect certain combinations of the linear sensors to map directly to the varying pointing model parameters.
We calculate and include these linear combinations as features.
Furthermore, we expect that the direction of the wind relative to the orientation of the telescope is more important than the absolute direction of the wind, so we also calculate the angular difference between the current telescope \aztext{} and the \aztext{} of oncoming wind.
We do not try engineering meaningful linear combinations of the thermometers.
We expect that doing so is an area for improving the accuracy and efficiency of future models.

\subsection{Dataset}
\label{sec:dataset}

We compile a dataset from years of archival data taken with the SPT-3G and EHT cameras.
The data taken with the SPT-3G receiver come from the typical SPT observations of the CMB, as described in Section~\ref{sec:spt_point}.
That is, we use the observations of sources that appear in the maps and that are used in the last step of offline pointing correction.
The data taken with the EHT receiver come from pointing observations during EHT observing campaigns, as described in Section~\ref{sec:eht_point}.
In addition to the source maps, these archival data also include sensor data logged multiple times a second, from which we can recover values for the feature variables at the time of each observation.

The final training dataset consists of 121,369 source observations.
All sources are used for training a model to estimate $\afour$, while only a subset 73,187 sources are used for training a model to estimate $\asix$.
This discrepancy is due to the fact that there are two historical \eltext{} encoder operating modes, each with different \eltext{} pointing behavior.
We currently only ever operate in one of these modes, so only historical observations made in this particular mode are used to train the model for $\asix$.

Observations made as part of the SPT-3G survey comprise 97.3\% and 95.6\% of observations in the $\afour$ and $\asix$ training data, respectively.
Typical SPT observations of the CMB occur throughout the vast majority of the year, so these observations are plentiful.
Throughout the cold austral winter, the SPT-3G survey historically focused on one region of the sky, the ``winter field,'' with repeated observations over an \eltext{} range of $42\degree < \el < 70\degree$.
During the warmer austral summer, the sun may contaminate the winter field, so the SPT-3G survey is focused on other regions of the sky, the ``summer fields,'' which include \eltext{} ranges as low as 28\degree.
In this work, we use sources observed in the winter and summer fields between 2019 February and 2023 September.
For each observation map, we measure the offsets of bright sources with S/N $>10$ as detailed in Section~\ref{sec:spt_point}, resulting in about 3-20 sources per observation.
We use the measured pointing offsets to derive values for the target variables $\afour$ and $\asix$ which minimize the pointing offset for each source, and we record averages of the archived feature data at a few time samples during which the center of the focal plane is pointed closest to the expected source location.

The SPT-3G observations are useful for training a machine learning model.
They are plentiful, and near constant observing ensures that we probe a broad range of weather condition feature space.
However, these data also present an interesting challenge.
The winter field covers a relatively small \eltext{} range, so we cannot expect it alone to train a model that will extrapolate well to other \eltext{} ranges.
The summer fields extend to lower \eltext{}, but this introduces a new problem since these elevations are only observed during the warm summer months.
Accurately pointing the SPT for EHT observations require predictions which work at low \eltext{}, and scheduled EHT observations typically occur in the austral fall after the sun sets and temperatures approach the cold winter temperatures.

Observations made with the EHT spectrometer comprise 2.7\% and 4.4\% of observations in the $\afour$ and $\asix$ training data, respectively.
These observations are typically made at two times of year: in the austral fall when the annual EHT observing runs are typically held and in the austral summer when telescope operators are trained for those runs.
The data used here include 1,232 sources in austral fall 2021, 707 sources in austral summer 2022-3, 1,172 sources in austral fall 2023, and 145 sources in austral summer 2023-4.
For each source, we measure the offsets from raster-scanning over the source with a spectrometer, as described in Section~\ref{sec:eht_point}.
For each installation of the EHT mirrors, we use the measured offsets to fit for values of $\afive$ and $\asix$ that we expect describe the difference between the SPT-3G and EHT pointing models; we use these values to describe the domain difference between data derived through EHT and SPT-3G observations.
We then use the measured offsets to derive values for the target variables $\afour$ and $\asix$ which fully account for the pointing offset of each source, we correct the $\asix$ value for expected differences between the expected EHT and SPT-3G pointing model, and we record averages of the archived feature data over the few minutes spent scanning the source.

The EHT source observations complement the SPT-3G source observations during training.
While the SPT-3G source observations cover a limited range in \eltext{}, the EHT source observations cover broad ranges in \aztext{} and \eltext{}, by design.
The EHT source observations are less plentiful, but we expect the broad \eltext{} range of the sources to help the models extrapolate outside the \eltext{} regimes at which SPT-3G sources constrain the model well.
The EHT source observations usually happen during austral summer and fall, so they probe a broad range of weather condition feature space.
Finally, because of the SPT-3G observation scanning strategy and large focal plane, the time at which we scan over a source is ill-defined; the SPT-3G feature data are averages over instances in time potentially spanning tens of minutes.
This implies that precisely measuring a state of conditions for the telescope and weather during an SPT-3G source observation is difficult to do.
EHT source observations take a smaller amount of time, all of which is dedicated to scanning over the source, so it is easier to describe the state of the telescope and weather at the time of observation.
All that said, EHT source observations come with difficulties, chiefly that the observations have lower S/N than the SPT-3G source observations, so statistical uncertainty in measuring a source position is higher.
Likewise, there are fewer visible sources, so the EHT data sample \eltext{} values less densely than SPT-3G data.

\begin{figure*}[!htb]
	\centering
	\includegraphics[width=0.7\linewidth]{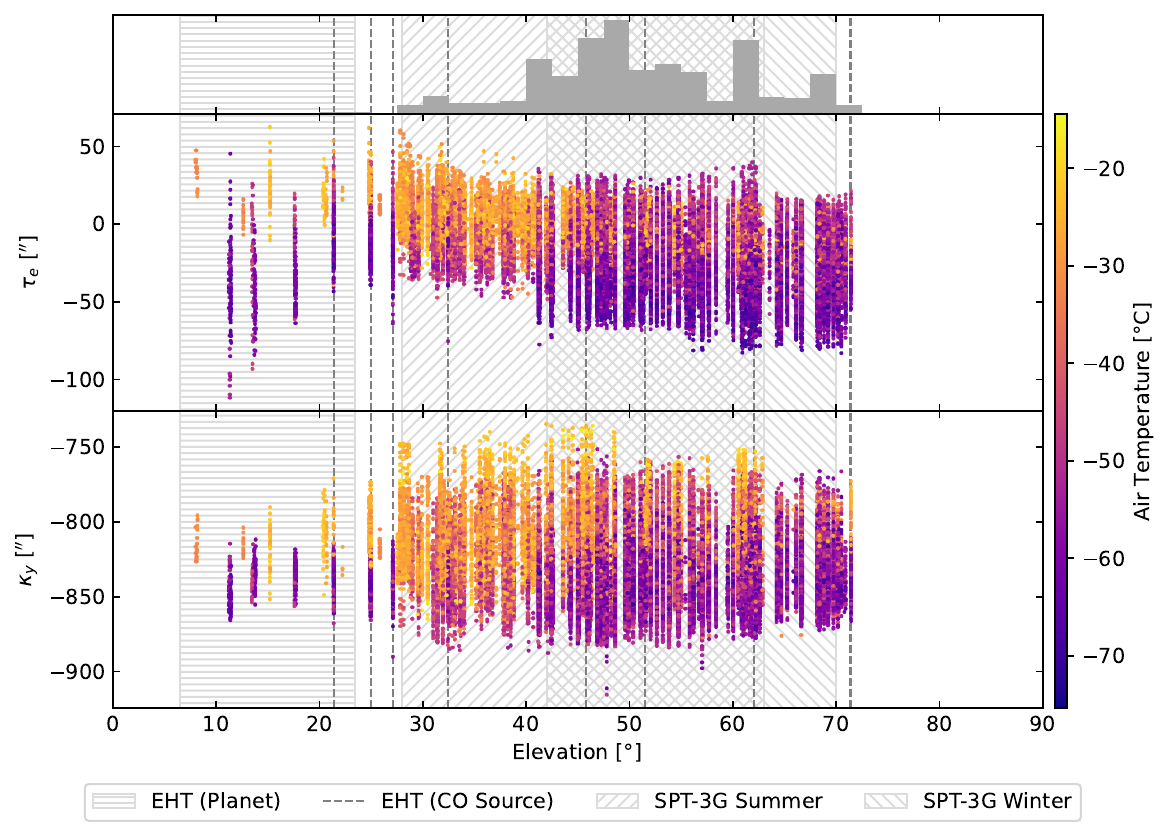}
	\caption{
		\small
		Target variables $\afour$ and $\asix$ in the training dataset to be learned by the machine learning model. 
		Sources with \eltext{} $42\degree < \el < 70\degree$ are predominantly from SPT-3G observations during the cold austral winter.
		Sources with \eltext{} $28\degree < \el < 42\degree$ are predominantly from SPT-3G observations during the warmer austral summer.
		Other sources observed with the EHT receiver are Other sources observed with the EHT receiver are either individual sources with prominent carbon monoxide (CO) emission lines, which appear at nearly constant \eltext{} at the geographic South Pole, or bright thermal sources like planets, which can appear over a range of \eltext{}.
	}
	\label{fig:dataset}
\end{figure*}

Figure~\ref{fig:dataset} shows the combined dataset used for training the ML model.
In the figure, one can see the difficulties in covering broad ranges of feature space by including both SPT-3G and EHT source observations.

While our dataset is very specific to our instruments and observing location, we understand that these data may be useful for other researchers developing their own methods.
The data used in this work are available upon request.

We split the data randomly into two sets: 80\% of the data were used for training and validation, and 20\% of the data were withheld as a final test set.

\subsection{Architecture Selection and Training}

Our scenario is a supervised regression problem with tabular data and two target variables: the pointing model parameters $\afour$ and $\asix$.
We utilize mean squared error as our loss metric.
Our chosen model architecture is XGBoost \cite{chen16}, a type of gradient-boosted decision tree that performs particularly well on tabular data \cite{shwartz22}.
In earlier stages of model development, we also explored using neural networks, but we found that these models performed worse than XGBoost models, especially when extrapolating to \eltext{} ranges outside of the training data.
XGBoost models learn only one target variable, so we aim to develop two models, one for each of the target variables.

By using five-fold cross validation, we tune the XGBoost hyperparameters \texttt{max\_depth}, which is the maximum depth of each tree; \texttt{early\_stop}, which is the number of training steps to perform without loss improvement before stopping training; and \texttt{eta}, which is the learning rate at which the boosting algorithm learns from each iteration.
We randomly split the 80\% of data apportioned for training into five ``folds.''
For each combination of hyperparameters, we train five XGBoost models, for which each fold serves once as a withheld test set.
During training, 20\% of the data in non-test folds are used as a validation set for early stopping.
We select the optimal combination of hyperparameters by the average performance on the test sets over all folds.
We utilize two rounds of cross validation: the first round exploring initial guesses for hyperparameter values and the second round exploring other values in directions that the training might prefer.
We list the hyperparameter values explored and ultimately selected in Tables~\ref{tab:cv_a4} and \ref{tab:cv_a6}.

\begin{wstable}
\centering
\caption{Hyperparameter Tuning for $\afour$}
\begin{tabular}{@{}lccc@{}}
\toprule
Hyper-					& Round 1  				& Round 2			& Final \\ 
parameter				& Values  				& Values 			& Selection \\
\colrule
\texttt{max\_depth}		& [2, 3, 4, 5]			& [4, 5, 6, 7]		& 5 \\
\texttt{early\_stop}	& [50, 100, 300]		& [100, 300, 500]	& 500 \\
\texttt{eta}			& [.03, .05, .1, .3]	& [.01, .03, .05]	& 0.05 \\
\botrule
\end{tabular}
\begin{flushleft}
{\small
A list of the XGBoost hyperparameters tuned during cross validation training the $\afour$ model.
The hyperparameter values explored during two rounds of cross validation are listed along with the values selected for training the final model.}
\end{flushleft}
\label{tab:cv_a4}
\end{wstable}

\begin{wstable}
\centering
\caption{Hyperparameter Tuning for $\asix$}
\begin{tabular}{@{}lccc@{}}
\toprule
Hyper-					& Round 1  				& Round 2			& Final \\ 
parameter				& Values  				& Values 			& Selection \\
\colrule
\texttt{max\_depth}		& [2, 3, 4, 5]			& [4, 5, 6]			& 4 \\
\texttt{early\_stop}	& [50, 100, 300]		& [100, 300, 500]	& 300 \\
\texttt{eta}			& [.03, .05, .1, .3]	& [.03, .05]		& 0.05 \\
\botrule
\end{tabular}
\begin{flushleft}
{\small
A list of the XGBoost hyperparameters tuned during cross validation training the $\asix$ model.
The hyperparameter values explored during two rounds of cross validation are listed along with the values selected for training the final model.}
\end{flushleft}
\label{tab:cv_a6}
\end{wstable}

We also experimented with excluding \eltext{} explicitly as a feature variable.
We did so to explore the fidelity of extrapolation during real-world operation to \eltext{}s outside of the training range, and we hypothesized that the model might not need \eltext{} to learn those relevant physical processes affecting pointing.
We performed the hyperparameter tuning process described above to train models for $\afour$ and $\asix$ both with and without \eltext{} as an input feature.
Including \eltext{} as a feature had significantly better RMSE on withheld validation data, and models with explicit \eltext{} dependence performed well on a variety of stress tests.
We selected the models which did include explicit \eltext{} dependence as our final models.

We trained the final models with the selected hyperparameter values.
The training data consisted of all data used during cross validation.
We use 20\% of the training data as a validation set for early stopping.
We evaluated the final models on the 20\% of data initially withheld as a test set.
The $\afour$ model achieved 3\farcs01 RMSE on the test set, which we estimate corresponds to a 2\farcs14 RMSE in \xeltext{}.
The $\asix$ model achieved 3\farcs57 RMSE on the test set, which is equivalent to the estimated RMSE in \eltext{}.
These errors achieve our goal of $<5\arcsec$ RMSE in \xeltext{} and in \eltext{}.

\subsection{Model Validation and Stress Testing}

To minimize risk to the telescope structure or receiver hardware, any changes to the control system must be vetted strongly beforehand.
To vet the ML models and ensure they perform as expected, it is critical to stress test them before deployment and to try interpreting what they have learned.
In addition to evaluating the model on a withheld test set, we performed several checks.

\begin{figure*}[!htb]
	\centering
	\includegraphics[width=0.8\linewidth]{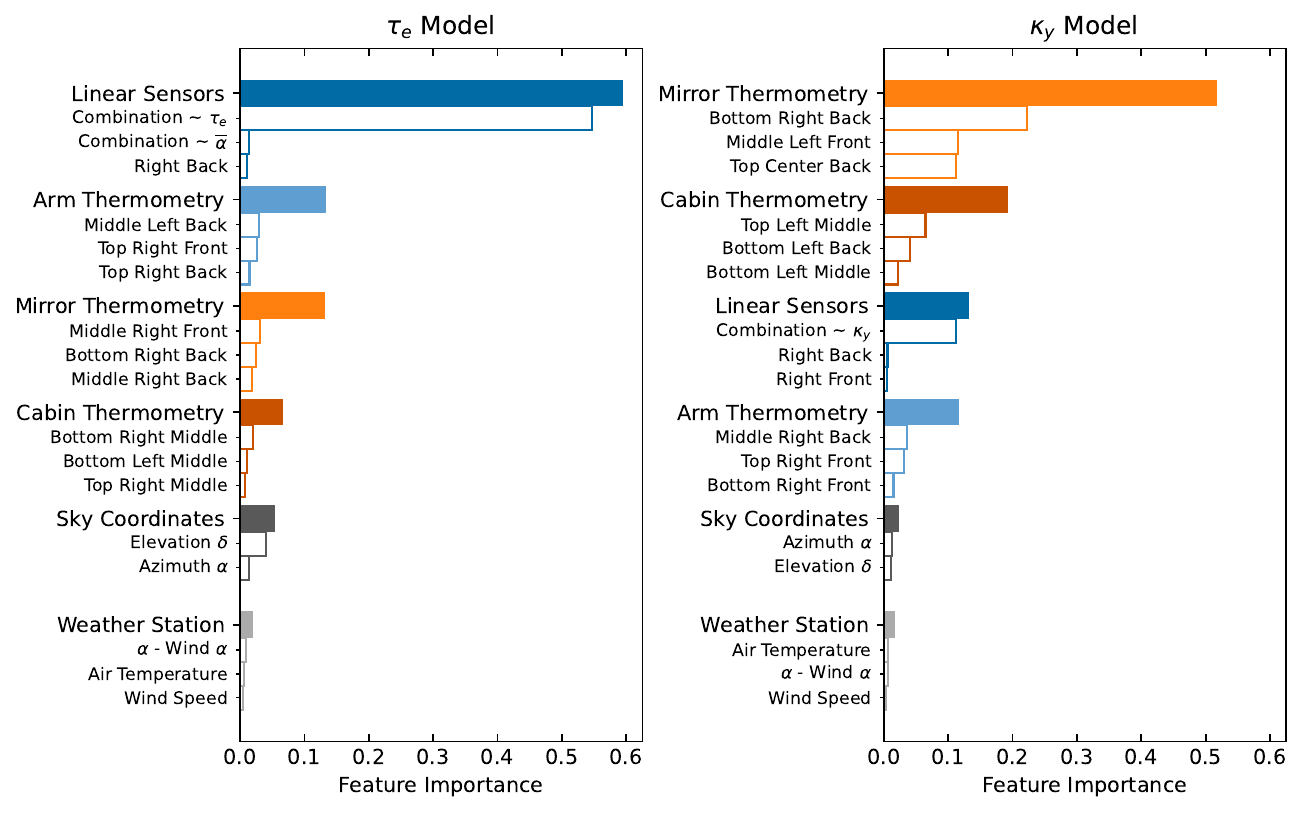}
	\caption{
		\small
		Feature importances for \textit{(left)} the $\afour$ model and \textit{(right)} the $\asix$ model.
		The solid bars show the total combined importance for each category of features, and the unfilled bars show the three most important features in each category.
	}
	\label{fig:feature_importance}
\end{figure*}

One check involved interpreting feature importance to investigate how the model made decisions.
For instance, previous exploration of the dataset supported the physical intuition that certain combinations of the linear sensors should correlate directly with $\afour$ and $\asix$.
We hoped that those linear sensor combinations would be crucial features as we expected that they explained much of, but not all of, the variation in our target variables.
Figure~\ref{fig:feature_importance} shows the importance as calculated by the XGBoost software for each category of feature and for the most important features in each category.
Indeed, linear sensor combinations ranked highly in the list of learned feature importance, especially for the $\afour$ model.
In addition, thermometry on the boom supporting the primary mirror was particularly important for the $\asix$ model, suggesting potential thermal deformation of this structure.
Features with low importance corresponded with features that physical intuition might expect to be secondary corrections.
For example, thermometers along the bottom of the yoke arms had low importance, as we expected them to capture physical processes measured more directly by the linear sensors.

Another stress test involved extrapolating the models beyond training data feature ranges, especially in \eltext{}.
We selected subsets of the data representative of standard observing conditions in the austral summers and winters and averaged the feature data for those time periods.
For average seasonal weather conditions, we evaluated the models over a grid in \aztext{} around the sky and in \eltext{} ranging from 2\degree{} to 88\degree.
We confirmed that the outputs did not diverge significantly, particularly for $\el < 10\degree$ and $\el > 72\degree$, ranges over which there are little to no training data.
We confirmed that the outputs interpolated somewhat smoothly between the training data within the range $10\degree < \el < 28\degree$, an \eltext{} range which training data sampled more sparsely.
We confirmed that expected important physical dependencies were present at extreme \eltext{} values; for a source at $\el = 3\degree$, we tested input values of the linear sensor combination features and observed a linear correlation with the outputs as expected by physical intuition.

A final stress test aimed to explore whether deployment might encounter a domain shift.
As described in Section~\ref{sec:dataset}, the vast majority of the training data was collected with the SPT-3G receiver during standard SPT operations.
Data taken with the EHT receiver composed a small fraction of the dataset, yet this was the domain for which we developed the models.
We separated the EHT data within the withheld test set and evaluated the models on just this subset of the test data.
The $\afour$ model achieved 7\farcs41 RMSE on the test set, which we estimate corresponds to a of 3\farcs54 RMSE in \xeltext{}.
The $\asix$ model achieved 5\farcs77 RMSE on the test set, which is equivalent to the estimated RMSE in \eltext{}.
These errors were worse than the errors on the entire test set, but they were within or near our goal.
We expected that the EHT observations had worse S/N than the SPT-3G observations, which added some uncertainty to these measurements and therefore a larger error in trained model predictions.
Thus, we assumed that these errors were within tolerance for this work.
We further discuss potential domain shifting in Section~\ref{sec:discussion}.

\section{Deployment and Results}
\label{sec:results}

Deployment consisted of three steps.
First, in austral summer 2023-4, we made the necessary changes to the telescope control system software to monitor incoming feature data, hand those data to the models, and apply the output values as adjustments to the baseline pointing model.
During this time, we performed test observations with preliminary models to ensure proper integration with the control system.
Second, as part of setup before the EHT observing campaign in 2024 April, we swapped out the preliminary models for the final trained models.
As described in Section~\ref{sec:eht_point} for previous EHT observing campaigns, the setup included pointing observations to fit for the proper baseline values of $\afive$ and $\asix$.
Third, throughout the remainder of the 2024 April EHT observing campaign, we performed a number of test observations with the model corrections active.
These data serve as \textit{in situ} evaluations of the accuracy of the trained models.
Additionally, we use the model corrections for all other observations during the observing campaign.
This section details the steps of the model deployment and describes the measured test data.

\subsection{Control System Integration}
\label{sec:integration}

The SPT is controlled by a software system called the Generic Control Program (GCP).
The GCP is written mostly in the C++ and C coding languages and is in use on a handful of other telescopes.
The GCP has the ability to perform a number of tasks, including controlling telescope movement, operating detectors, and collating and storing data.
The GCP consists of three main components: the Control layer, which provides control over the whole experiment; the Antenna layer, which provides an interface between the GCP and two computers which drive the telescope; and the Mediator layer, which mediates between the Control layer and all other hardware control processes, including data sources and the Antenna layer.
For more details on the SPT control system including the GCP, see \citet{story12}.

Typical observations are run via ``schedules,'' scripts written in a GCP-specific scripting language.
The schedules consist of sequential commands to the Control layer, which instructs the requested actions of the other GCP components.
For instance, EHT pointing observations are conducted with a schedule command that includes the following instructions: update the target \aztext{} and \eltext{} to match the expected source position, wait for the telescope to slew to the target location, begin recording data, scan the telescope over the source, then stop recording data.  

Prior to running an observing schedule, the telescope must be set up.
Typical setup includes defining values for pointing model parameters, which are handed as arguments to the GCP.
When given a target \aztext{} and \eltext{}, the GCP will compute the same corrections to encoder \aztext{} and \eltext{} as described in Section~\ref{sec:pointing_model} using the pointing model parameter values provided during setup.

In austral summer 2023-4, we added new commands and variables to the GCP.
First, we added a set of pointing model parameter ``adjustments'' as new variables.
When the GCP is asked to move the telescope to a target \aztext{} and \eltext{}, it now evaluates the pointing model with parameter values equal to the sum of values provided during setup and values stored as adjustments.
Second, we added a command that can update the stored adjustments while within an observation schedule.
One could thus provide a best guess at average pointing model parameters during setup and then update only relatively small adjustments during actual operation.
Finally, we added a command that uses the ML models to translate current weather conditions and telescope status into pointing model parameter adjustments.

The last command above uses the ML models as follows:
The Control layer runs the observation schedule.
First, prior to observing a target source, the Control layer commands the Antenna layer to slew the telescope to the source and track the source as it moves on the sky.
Second, the Control layer runs a Python script that loads the trained ML models into memory and awaits input data.
Next, the Control layer collects the feature data sampled once per second from the Mediator layer and supplies these as inputs to the ML models in the Python script.
Optionally, these feature data can be averaged over a specified length of time to reduce instrument noise and weather variability before an output is returned; we use a default averaging time of $10\,$s, a period that balances feature variation with time constraints in an analysis of some historical observations.
Finally, the Control layer updates the pointing model parameter adjustments using the outputs from the ML model as returned by the Python script.
The Antenna layer is instructed to update the telescope pointing based on the new adjustments.
In this way, we are able to utilize the state of the weather and telescope structure in the same conditions as the impending source observation.
We scheduled this command to occur before every science target source observation during the EHT observing run.

\subsection{Baseline Pointing Model}
\label{sec:baseline}

Setup for the EHT observing run began on 2024 March 29, a few days before the first possible observing session on 2024 April 3.
After the receiver was set up for observing, we began pointing observations.
The purpose of these observations was to fit for the baseline pointing model parameter values, i.e. the values about which the ML adjustments would fluctuate.
We must perform this exercise and cannot assume historical averages because we expect the values of $\afive$ and $\asix$ to vary between installations of the EHT receiver and mirrors as described in Section~\ref{sec:eht_point}.

We continued these pointing observations for most of three days and stopped prior to the first observing session in which the SPT participated on 2024 April 6. 
This dataset included 890 source observations with ML adjustments active during observation.
With this dataset, we fit for optimal baseline values of $\afive$ and $\asix$ which incorporate the difference between SPT-3G and EHT operating modes.
These values were $4\arcmin41\farcs85$ and $-12\arcmin30\farcs78$, respectively.
We use $\afour = -16\farcs65$ based on an analysis of historical data as we do not expect this to differ between operating modes or EHT installations.
We measure $\atwo$ and $\athree$ directly as described in Section~\ref{sec:var_point} at the start of each day.
We use the typical values of $\azero$, $\aone$, $\aseven$, and $\azzero$.
This set of parameter values is used for the rest of the observing run.

We do not use these data to test our model \textit{in situ} directly, as described in Section~\ref{sec:in_situ}.
While it is possible to do so, we would like to know how the ML models perform once the entire system is set up with the final baseline parameter values.
Excluding these data from the \textit{in situ} model testing ensures a dataset clean of potential confounding factors introduced by a set of different pointing model parameters values used during observing.

Using these data, we also fit for the constant values of $\afour$, $\afive$, and $\asix$ that correspond to operating without ML adjustments active.
That is, these are the pointing model parameter values we would hold fixed for the entirety of observations if we conducted our pointing in the way we historically have as described in Section~\ref{sec:eht_point}.
These values are $\afour = -13\farcs62$, $\afive = 4\arcmin40\farcs70$, and $\asix = -12\arcmin39\farcs57$.
We use these values in Section~\ref{sec:in_situ} to estimate what the residual pointing errors would be without the ML adjustments active.

\subsection{In Situ Test Observations}
\label{sec:in_situ}

After setup was complete, we spent the remaining time of the scheduled 2024 April EHT observing campaign either performing the scheduled EHT observations or conducting additional pointing observations.
We use the source observations conducted during all those activities to compile a dataset used to test the ML models \textit{in situ}.

This dataset is composed of 1,173 source observations between 2024 April 7 and 2024 April 14.
We spent a comparatively large amount of time observing Mars and Saturn because these sources were at very low elevation and we hope to reuse these data for training future models.
For each source, we measure the observed pointing error in the usual way.
The size of these errors represent the remaining error in the estimation power of the ML models.

We estimate the accuracy gained from using the ML models by estimating what the errors would have been had we observed without using the ML models.
We calculate these hypothetical errors by taking the observed offsets with ML adjustments active, using the pointing model to calculate the ideal \aztext{} and \eltext{} encoder positions that would have given perfect pointing, then calculating the inverse of the pointing model using the static baseline pointing model parameter values listed in Section~\ref{sec:baseline}.

\begin{figure*}[ht!]
	\centering
	\includegraphics[width=1\linewidth]{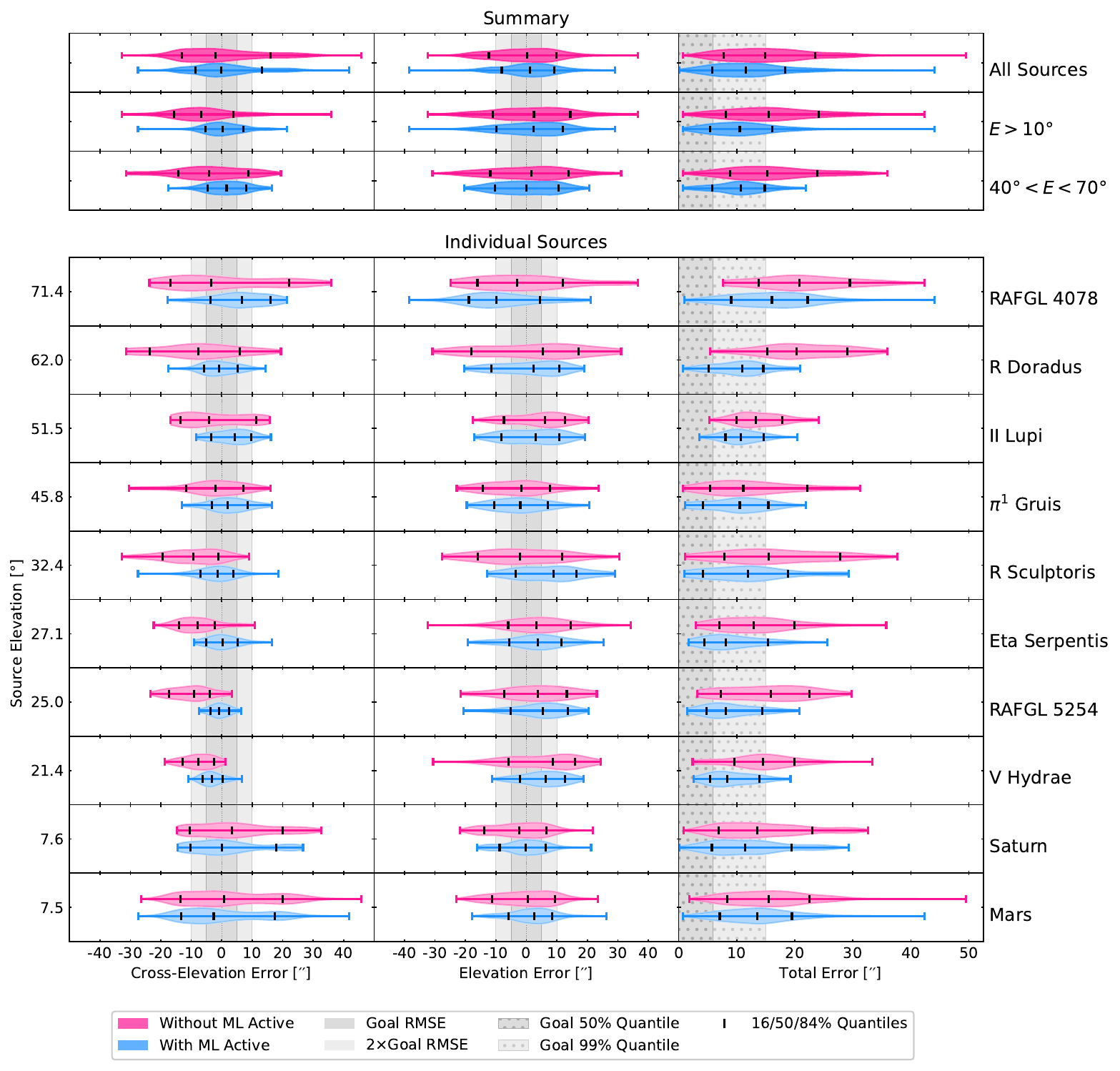}
	\caption{
		\small
		Errors in source position as measured with the trained models deployed and active on the telescope during the 2024 April EHT observations.
		The errors without ML active are estimations calculated using the pointing model equations assuming static values for $\afour$ and $\asix$.
	}
	\label{fig:offsets}
\end{figure*}

A breakdown of the number of observations for each source and their residual pointing errors is listed in Table~\ref{tab:val_srcs} and visualized in Figure~\ref{fig:offsets}.
The table and figure also include summary statistics for all sources combined, sources with $\el > 10\degree$ (i.e. sources well within the \eltext{} range of training data), and sources with $40\degree < \el < 70\degree$ (i.e. sources within the \eltext{} range of the SPT winter field where there is the most training data).

For all sources, using the ML adjustments resulted in a reduction of errors in \xeltext{} from $0\farcs2\pm13\farcs5$ to $1\farcs3\pm10\farcs5$ and in \eltext{} from $-0\farcs5\pm10\farcs9$ to $0\farcs9\pm8\farcs7$, where we report mean offset and the standard deviation scatter about that mean.
These modest improvements fall shy of our goal of $0\arcsec\pm5\arcsec$ along each of those axes.
Using the ML adjustments resulted in a 22\% reduction in the average error measured along a great circle, from $15\farcs6$ to $12\farcs2$.
For sources with $40\degree < \el < 70\degree$, using the ML adjustments resulted in a reduction of errors in \xeltext{} from $-4\farcs2\pm11\farcs6$ to $1\farcs8\pm6\farcs4$ and in \eltext{} from $1\farcs0\pm12\farcs5$ to $0\farcs1\pm9\farcs3$.
These improvements for sources in this \eltext{} range also fall shy of our goal of $0\arcsec\pm5\arcsec$ along each of those axes, but they show more improvement than all sources.
Using the ML adjustments resulted in a 33\% reduction in the average combined error measured along a great circle, from $15\farcs9$ to $10\farcs6$.
In general, errors in \aztext{} showed a greater improvement than errors in \eltext{}.

\begin{table*}[ht!]
\scriptsize
\centering
\caption{\textit{In Situ} Test Observations}
\begin{tabular}{lccccccccc}
\toprule\\[-6pt]
				& 				& 			& 		& \multicolumn{3}{c}{Errors Without ML Adjustments [\arcsec]}	& \multicolumn{3}{c}{Errors With ML Adjustments [\arcsec]} \\
Source Name		& \elmath{} [\degree]	& \# Obs.	& Rel. S/N	& \Xeltext{}	& \Eltext{}	& Total				& \Xeltext{}	& \Eltext{}	& Total \\
\colrule
All Sources		& 0---90		& 1173 		& --- 	& $0.2\pm13.5$	& $-0.5\pm10.9$	& $15.6\pm7.6$	& $1.3\pm10.5$	& $0.9\pm8.7$	& $12.2\pm6.4$ \\
All Sources		& 10---90		& 445 		& --- 	& $-5.8\pm11.1$	& $1.7\pm12.8$	& $16.2\pm7.8$	& $0.8\pm6.6$	& $1.7\pm10.4$	& $11.0\pm5.9$ \\
All Sources		& 40---70		& 166 		& --- 	& $-4.2\pm11.6$	& $1.0\pm12.5$	& $15.9\pm7.5$	& $1.8\pm6.4$	& $0.1\pm9.3$	& $10.6\pm4.4$ \\
\\[-6pt] \Hline \\[-6pt]
RAFGL 4078		& 71.4			& 54 		& 1.1 	& $1.0\pm16.9$	& $-0.2\pm15.6$	& $21.5\pm8.2$	& $5.8\pm9.2$	& $-7.8\pm12.0$	& $16.4\pm7.5$ \\
R Doradus		& 62.0			& 52 		& 1.1 	& $-7.8\pm13.3$	& $2.0\pm16.2$	& $21.3\pm7.1$	& $-0.8\pm6.0$	& $0.8\pm9.9$	& $10.7\pm4.6$\\
II Lupi			& 51.5			& 55 		& 1.9 	& $-1.8\pm10.4$	& $4.2\pm9.1$	& $14.0\pm4.2$	& $3.4\pm5.9$	& $1.6\pm8.9$	& $10.9\pm3.4$\\
$\pi^1$ Gruis	& 45.8			& 59 		& 1.3 	& $-3.1\pm10.3$	& $-3.0\pm10.3$	& $13.0\pm7.8$	& $2.5\pm6.5$	& $-1.9\pm8.8$	& $10.2\pm5.0$\\
R Sculptoris	& 32.4			& 56 		& 1.0 	& $-9.8\pm9.3$	& $-1.3\pm13.5$	& $17.0\pm8.9$	& $-1.7\pm6.8$	& $7.1\pm10.4$	& $12.3\pm7.4$\\
Eta Serpentis	& 27.1			& 59 		& 1.1 	& $-7.5\pm6.9$	& $3.8\pm11.1$	& $13.9\pm6.9$	& $0.6\pm5.3$	& $3.1\pm9.1$	& $9.6\pm5.4$\\
RAFGL 5254		& 25.0			& 55 		& 1.6 	& $-10.2\pm6.4$	& $3.4\pm11.2$	& $15.2\pm7.1$	& $-0.6\pm3.3$	& $4.4\pm8.6$	& $9.1\pm4.7$\\
V Hydrae		& 21.4			& 55 		& 1.6 	& $-7.6\pm4.9$	& $5.2\pm11.6$	& $14.3\pm6.0$	& $-3.0\pm3.5$	& $5.8\pm7.0$	& $9.3\pm4.0$\\
Saturn			& 7.6			& 392 		& 7.5 	& $4.7\pm12.8$	& $-2.8\pm8.9$	& $14.7\pm7.5$	& $2.3\pm11.5$	& $-0.6\pm7.3$	& $12.2\pm6.6$\\
Mars			& 7.5			& 336 		& 1.7 	& $3.1\pm14.3$	& $-0.7\pm9.5$	& $15.9\pm7.2$	& $0.8\pm13.0$	& $1.8\pm7.3$	& $13.6\pm6.4$\\
\botrule
\end{tabular}
\begin{flushleft}
{\footnotesize
Errors in source position as measured with the trained models deployed and active on the telescope during the 2024 April EHT observations.
The errors without ML active are estimations calculated using the pointing model equations assuming static values for $\afour$ and $\asix$.
S/N for each source is approximated relative to the source detected with the lowest average S/N.
Simulations of our source-fitting procedure indicate that any source with relative S/N $\lesssim 1.3$ may have even lower S/N, and these sources may have noise contributions to measured \xeltext{} and \eltext{} offsets on the order of the total measured offset distributions.
}
\end{flushleft}
\label{tab:val_srcs}
\end{table*}

\section{Discussion}
\label{sec:discussion}

Our goal defined in Section~\ref{sec:eht_point} was to achieve residual RMSE of 5\arcsec{} in \xeltext{} and 5\arcsec{} in \eltext{}, corresponding to a median error of 5\farcs9 and mean error of 6\farcs3 along a great circle.
Although we did not meet this goal, we made significant progress towards those ends and demonstrated a proof of concept for improvement in the future.

\begin{figure}[!ht]
	\centering
	\includegraphics[width=1\linewidth]{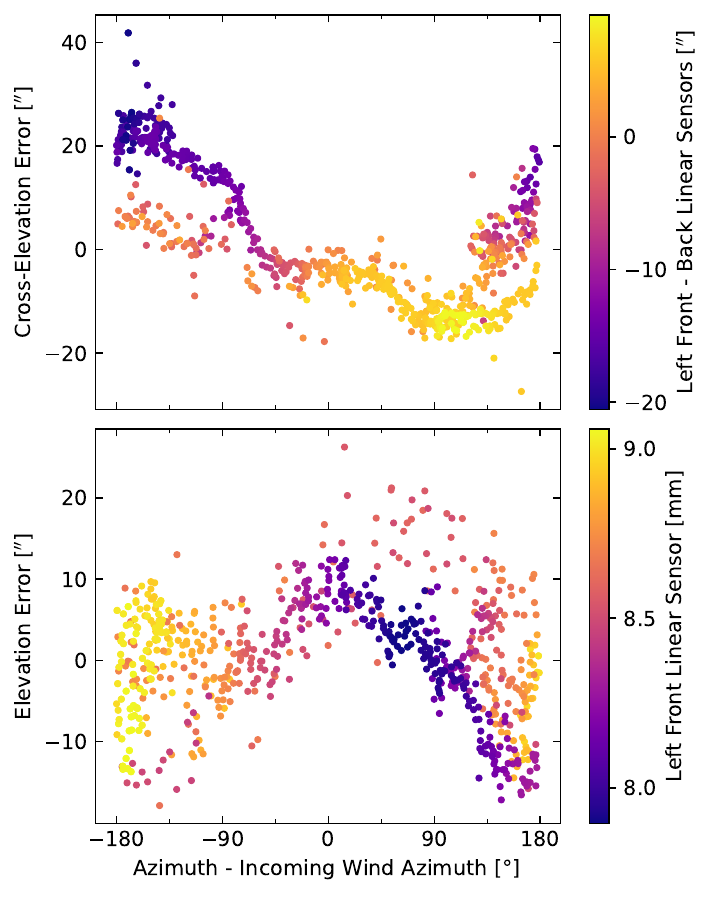}
	\caption{
		\small
		Measured pointing offsets for Saturn and Mars during \textit{in situ} test observations with ML corrections active.
		Remaining \xeltext{} errors \textit{(top)} can be described by a function of features such as \aztext{} and a combination of the linear sensors.
		Remaining \eltext{} errors \textit{(bottom)} can be described by a function of features such as \aztext{} and a combination of the linear sensors.
		The functions can be learned by future ML models with training on these data.
	}
	\label{fig:saturn_mars}
\end{figure}

A cursory look at errors for all sources suggests only modest improvement.
However, 62\% of all data were of Saturn and Mars.
These sources were at low \eltext{}s that required extrapolation outside the training range, so we should not expect the models to perform as well for these sources.
Nevertheless, the errors in these observations do demonstrate some improvement.
This suggests that the models have learned some physical dependencies that are applicable generally.
Furthermore, we investigated whether the residual errors for these observations correlated with any of the feature variables.
We see that the errors are not distributed purely randomly; they depend on the sky locations of the sources and on the weather conditions.
Examples of the error dependence on some select feature variables are shown in Figure~\ref{fig:saturn_mars}.
The remaining \xeltext{} and \eltext{} errors can be described by functions of input features that can be learned by future ML models with training on these data.
Thus, these observations of Saturn and Mars will help future models extrapolate to lower elevations.

A closer look at errors for sources within the $\el > 10\degree$ training range shows greater improvement than for all sources.
This improvement is particularly pronounced for \aztext{} errors as a result of the $\afour$ model.
This improvement is even more pronounced for sources within the highly populated $40\degree < \el < 70\degree$ training range.
For those sources, the standard deviation of errors in \xeltext{} was reduced significantly from 11\farcs6 to 6\farcs4, nearly our goal, and the full width of the distribution of \eltext{} errors including outliers  was reduced by about 20\arcsec.
This resulted in an improvement in the average combined pointing error of 33\%.

Our target goal was derived based on loss of gain to a typical source given our model for the 345~GHz EHT beam, a two-dimensional Gaussian with 30\arcsec{} FWHM.
Alternatively, we could define other benchmarks for evaluating the effectiveness of our models.
Assuming that we meet our stated goal and that pointing errors along a great circle are Rayleigh-distributed with median error of 5\farcs9, then only 1.1\% of sources would have a pointing error worse than 15\arcsec, corresponding to a loss of gain more than 50\%.
We use this as yet another evaluation metric for our \textit{in situ} test observations.
We estimate that without ML corrections, 52.4\% of sources with $40\degree < \el < 70\degree$ have pointing offsets worse than this 15\arcsec{} benchmark.
With our ML corrections active, only 13.9\% of sources with $40\degree < \el < 70\degree$ have pointing offsets worse than this benchmark.
This serves as a proof of concept for the effectiveness of our models with sufficient training.

An even closer look at individual sources shows some errors are already within our stated goal.
In particular, the \xeltext{} errors for the sources R Doradus, R Sculptoris, Eta Serpentis, RAFGL 5254, and V Hydrae are all at or near our goal.
Furthermore, the sources besides Saturn and Mars with the largest outliers in total error are RAFGL 4078, R Sculptoris, and Eta Serpentis; these sources also have lower relative S/N than the other sources, suggesting that the magnitude of some of these errors are due to statistical uncertainty while fitting for location.

Despite the demonstrated benefit of using our trained models, we must address remaining curiosities to develop better models in the future.
First, we do not fully understand how the models generalize from the combined SPT-EHT training data to the specific EHT use case.
Training RMSE is worse for the EHT sources in withheld test data than for all sources.
\textit{In situ} test observation errors with the EHT receiver are worse than those in the withheld test set.
Second, we do not fully understand the performances of the two trained models.
For the $\afour$ model, the \aztext{} errors for some sources are within our goal while others are not.
Between the $\afour$ and $\asix$ models, the \eltext{} errors show less improvement in general than the \aztext{} errors.
For all these curiosities, we will posit some potential causes and solutions.

Some of these curiosities could be explained by differences between the domains of SPT-3G and EHT data.
If there are differences between the domains, model performance would be impacted by the following:
First, the quantity of SPT-3G training data far surpasses EHT training data, so the models are trained predominantly to learn physical behavior in the SPT-3G domain.
Second, EHT observations have lower S/N and thus noisier measured sky positions, so the underlying physical dependence on feature variables is obscured by more noise and is harder to learn.
Third, SPT-3G data cover a smaller \eltext{} range than EHT data, and so underlying physical dependencies learned over the SPT-3G \eltext{} range may not extrapolate well to the EHT range.

Potential domain differences between the SPT-3G and the EHT operating modes could include the following:
First, we may be making incorrect assumptions about the differences between the SPT-3G and EHT pointing models.
We convert the EHT data into a form similar to the SPT-3G data for training as described in Section~\ref{sec:dataset}.
That procedure assumes that the only differences are constant offsets in $\afive$ and $\asix$.
If this assumption is incorrect, we may introduce a bias to EHT training data not present in SPT-3G data or in real EHT performance.
Furthermore, the training data are gathered differently for SPT-3G and EHT.
During SPT-3G observations, it is difficult to define the time at which the telescope scanned over a source and the weather conditions at that time because the SPT-3G observations involve scanning over the source multiple times with a large field of view.
Thus, the SPT-3G feature data are averages over instances in time potentially spanning tens of minutes.
During EHT observations, we make direct source observations with clearly defined feature values.

Some of the curiosities could also be explained by our implementation method described in Section~\ref{sec:integration}.
The models are trained on data measured during an observation.
However, we make $\afour$ and $\asix$ predictions based on weather conditions just before a source observation.
If the feature variables change significantly between these two times, the ideal values of $\afour$ and $\asix$ would change and introduce pointing errors.

Most all of these questions can be solved with more training data.
In particular, we look forward to SPT-3G data taken at broader \eltext{} ranges, including the recently completed SPT Wide survey \cite{prabhu24}.
This survey includes sources covering \eltext{} $20\degree < \el < 80\degree$ with many high \eltext{} sources observed during the warm austral summer and many low \eltext{} sources observed during the cold austral winter.
These data will enable a more careful investigation of potential domain differences.
We also look forward to more EHT data.
These data will improve all aspects of training future models for use with EHT observations.
Finally, we look forward to exploring other training methods.
For instance, we could attempt weighting EHT data points higher.
We could explore pre-training the models in different phases, such as pre-training without \eltext{} as a feature before reintroducing it, or pre-training on SPT-3G data alone before reintroducing EHT data.

\section{Conclusion}
We demonstrate that ML can be used to improve the real-time pointing accuracy of the SPT.
We trained two models, one for correcting \xeltext{} errors and one for correcting \eltext{} errors, on historical observations of astronomical sources.
The models generalize well to withheld test data, achieving RMSE of 2\farcs14 in \xeltext{} and 3\farcs57 in \eltext{}, both below our goal of 5\arcsec.
We deployed these models on the telescope and made \textit{in situ} test observations during the 2024 April EHT observing campaign to test the accuracy of these models in the real world.

As a proof of concept, the models performed well for sources with \eltext{} $40\degree < \el < 70\degree$, an \eltext{} range within which there were the most training data.
With the use of our models active on the telescope, the standard deviation of errors in \xeltext{} improved from 11\farcs6 to 6\farcs4, and the full width of the distribution of \eltext{} errors including outliers was reduced by about 20\arcsec.
The improvement in the average combined pointing error is 33\%: from 15\farcs9 to 10\farcs6.
With sufficient training data, we expect future models will perform well across all elevations.

Upcoming datasets taken will improve training for future models.
For instance, the recently completed SPT Wide survey includes sources covering \eltext{} $20\degree < \el < 80\degree$.
These data will fill out significant ranges of feature space currently unconstrained by SPT-3G data.
In addition, continued observations with the EHT receiver, including those made as part of this work, can be used for future training.

This work demonstrates a significant improvement in the pointing accuracy of the SPT, and it will enable the expansion of science possible with the SPT as part of the EHT collaboration.

\section*{Acknowledgement}
The South Pole Telescope program is supported by the National Science Foundation (NSF) through awards OPP-1852617 and OPP-2332483.
Partial support is also provided by the Kavli Institute of Cosmological Physics at the University of Chicago. 
Argonne National Laboratory’s work was supported by the U.S. Department of Energy, Office of High Energy Physics, under contract DE-AC02-06CH11357. 
The UC Davis group acknowledges support from Michael and Ester Vaida. 
Work at Fermi National Accelerator Laboratory, a DOE-OS, HEP User Facility managed by the Fermi Research Alliance, LLC, was supported under Contract No. DE-AC02-07CH11359. 
The Melbourne authors acknowledge support from the Australian Research Council’s Discovery Project scheme (No. DP210102386). 
The Paris group has received funding from the European Research Council (ERC) under the European Union’s Horizon 2020 research and innovation program (grant agreement No 101001897), and funding from the Centre National d’Etudes Spatiales. 
The SLAC group is supported in part by the Department of Energy at SLAC National Accelerator Laboratory, under contract DE-AC02-76SF00515.
PMC, TMC, AEL, and DPM acknowledge support from NSF award AST-2034306.
Preprint of an article submitted for consideration in the Journal of Astronomical Instrumentation © 2024 [copyright World Scientific Publishing Company] [\url{https://www.worldscientific.com/worldscinet/jai}].

\section*{ORCID}
{\small
P.~M.~Chichura \orcidlink{0000-0002-5397-9035} https://orcid.org/0000-0002-5397-9035\\
A.~Rahlin \orcidlink{0000-0003-3953-1776} https://orcid.org/0000-0003-3953-1776\\
A.~J.~Anderson \orcidlink{0000-0002-4435-4623} https://orcid.org/0000-0002-4435-4623\\
M.~Archipley \orcidlink{0000-0002-0517-9842} https://orcid.org/0000-0002-0517-9842\\
L.~Balkenhol \orcidlink{0000-0001-6899-1873} https://orcid.org/0000-0001-6899-1873\\
A.~N.~Bender \orcidlink{0000-0001-5868-0748} https://orcid.org/0000-0001-5868-0748\\
B.~A.~Benson \orcidlink{0000-0002-5108-6823} https://orcid.org/0000-0002-5108-6823\\
F.~Bianchini \orcidlink{0000-0003-4847-3483} https://orcid.org/0000-0003-4847-3483\\
L.~E.~Bleem \orcidlink{0000-0001-7665-5079} https://orcid.org/0000-0001-7665-5079\\
F.~R.~Bouchet \orcidlink{0000-0002-8051-2924} https://orcid.org/0000-0002-8051-2924\\
E.~Camphuis \orcidlink{0000-0003-3483-8461} https://orcid.org/0000-0003-3483-8461\\
T.-L.~Chou \orcidlink{0000-0002-3091-8790} https://orcid.org/0000-0002-3091-8790\\
T.~M.~Crawford \orcidlink{0000-0001-9000-5013} https://orcid.org/0000-0001-9000-5013\\
C.~Daley \orcidlink{0000-0002-3760-2086} https://orcid.org/0000-0002-3760-2086\\
D.~Dutcher \orcidlink{0000-0002-9962-2058} https://orcid.org/0000-0002-9962-2058\\
K.~R.~Ferguson \orcidlink{0000-0002-4928-8813} https://orcid.org/0000-0002-4928-8813\\
A.~Foster \orcidlink{0000-0002-7145-1824} https://orcid.org/0000-0002-7145-1824\\
R.~Gualtieri \orcidlink{0000-0003-4245-2315} https://orcid.org/0000-0003-4245-2315\\
G.~P.~Holder \orcidlink{0000-0002-0463-6394} https://orcid.org/0000-0002-0463-6394\\
J.~Kim \orcidlink{0000-0002-4274-9373} https://orcid.org/0000-0002-4274-9373\\
A.~E.~Lowitz \orcidlink{0000-0002-4747-4276} https://orcid.org/0000-0002-4747-4276\\
D.~P.~Marrone \orcidlink{0000-0002-2367-1080} https://orcid.org/0000-0002-2367-1080\\
M.~Millea \orcidlink{0000-0001-7317-0551} https://orcid.org/0000-0001-7317-0551\\
G.~I.~Noble \orcidlink{0000-0002-5254-243X} https://orcid.org/0000-0002-5254-243X\\
Z.~Pan \orcidlink{0000-0002-6164-9861} https://orcid.org/0000-0002-6164-9861\\
K.~A.~Phadke \orcidlink{0000-0001-7946-557X} https://orcid.org/0000-0001-7946-557X\\
C.~L.~Reichardt \orcidlink{0000-0003-2226-9169} https://orcid.org/0000-0003-2226-9169\\
J.~A.~Sobrin \orcidlink{0000-0001-6155-5315} https://orcid.org/0000-0001-6155-5315\\
C.~Umilta \orcidlink{0000-0002-6805-6188} https://orcid.org/0000-0002-6805-6188\\
N.~Whitehorn \orcidlink{0000-0002-3157-0407} https://orcid.org/0000-0002-3157-0407\\
W.~L.~K.~Wu \orcidlink{0000-0001-5411-6920} https://orcid.org/0000-0001-5411-6920\\
}

\bibliography{bib}

\end{document}

%% file: authors.tex
\author{
P.~M.~Chichura\,\orcidlink{0000-0002-5397-9035}\textsuperscript{1,2,*}, 
A.~Rahlin\,\orcidlink{0000-0003-3953-1776}\textsuperscript{3,2}, 
A.~J.~Anderson\,\orcidlink{0000-0002-4435-4623}\textsuperscript{4,2,3}, 
B.~Ansarinejad\textsuperscript{5}, 
M.~Archipley\,\orcidlink{0000-0002-0517-9842}\textsuperscript{2,3}, 
L.~Balkenhol\,\orcidlink{0000-0001-6899-1873}\textsuperscript{6}, 
K.~Benabed\textsuperscript{6}, 
A.~N.~Bender\,\orcidlink{0000-0001-5868-0748}\textsuperscript{7,2,3}, 
B.~A.~Benson\,\orcidlink{0000-0002-5108-6823}\textsuperscript{4,2,3}, 
F.~Bianchini\,\orcidlink{0000-0003-4847-3483}\textsuperscript{8,9,10}, 
L.~E.~Bleem\,\orcidlink{0000-0001-7665-5079}\textsuperscript{7,2}, 
F.~R.~Bouchet\,\orcidlink{0000-0002-8051-2924}\textsuperscript{6}, 
L.~Bryant\textsuperscript{11}, 
E.~Camphuis\,\orcidlink{0000-0003-3483-8461}\textsuperscript{6}, 
J.~E.~Carlstrom\textsuperscript{2,11,1,7,3}, 
C.~L.~Chang\textsuperscript{7,2,3}, 
P.~Chaubal\textsuperscript{5}, 
A.~Chokshi\textsuperscript{12}, 
T.-L.~Chou\,\orcidlink{0000-0002-3091-8790}\textsuperscript{3,2}, 
A.~Coerver\textsuperscript{13}, 
T.~M.~Crawford\,\orcidlink{0000-0001-9000-5013}\textsuperscript{2,3}, 
C.~Daley\,\orcidlink{0000-0002-3760-2086}\textsuperscript{14,15}, 
T.~de~Haan\textsuperscript{16}, 
K.~R.~Dibert\textsuperscript{3,2}, 
M.~A.~Dobbs\textsuperscript{17,18}, 
M.~Doohan\textsuperscript{5}, 
A.~Doussot\textsuperscript{6}, 
D.~Dutcher\,\orcidlink{0000-0002-9962-2058}\textsuperscript{19}, 
W.~Everett\textsuperscript{20}, 
C.~Feng\textsuperscript{21}, 
K.~R.~Ferguson\,\orcidlink{0000-0002-4928-8813}\textsuperscript{22,23}, 
K.~Fichman\textsuperscript{1,2}, 
A.~Foster\,\orcidlink{0000-0002-7145-1824}\textsuperscript{19}, 
S.~Galli\textsuperscript{6}, 
A.~E.~Gambrel\textsuperscript{2}, 
R.~W.~Gardner\textsuperscript{11}, 
F.~Ge\textsuperscript{24}, 
N.~Goeckner-Wald\textsuperscript{9,8}, 
R.~Gualtieri\,\orcidlink{0000-0003-4245-2315}\textsuperscript{25}, 
F.~Guidi\textsuperscript{6}, 
S.~Guns\textsuperscript{13}, 
N.~W.~Halverson\textsuperscript{26,27}, 
E.~Hivon\textsuperscript{6}, 
G.~P.~Holder\,\orcidlink{0000-0002-0463-6394}\textsuperscript{21}, 
W.~L.~Holzapfel\textsuperscript{13}, 
J.~C.~Hood\textsuperscript{2}, 
A.~Hryciuk\textsuperscript{1,2}, 
N.~Huang\textsuperscript{13}, 
F.~K\'eruzor\'e\textsuperscript{7}, 
A.~R.~Khalife\textsuperscript{6}, 
J.~Kim\,\orcidlink{0000-0002-4274-9373}\textsuperscript{28}, 
L.~Knox\textsuperscript{24}, 
M.~Korman\textsuperscript{29}, 
K.~Kornoelje\textsuperscript{3,2}, 
C.-L.~Kuo\textsuperscript{8,9,10}, 
K.~Levy\textsuperscript{5}, 
A.~E.~Lowitz\,\orcidlink{0000-0002-4747-4276}\textsuperscript{2,30}, 
C.~Lu\textsuperscript{21}, 
A.~Maniyar\textsuperscript{8,9,10}, 
D.~P.~Marrone\,\orcidlink{0000-0002-2367-1080}\textsuperscript{30}, 
E.~S.~Martsen\textsuperscript{3,2}, 
F.~Menanteau\textsuperscript{15,31}, 
M.~Millea\,\orcidlink{0000-0001-7317-0551}\textsuperscript{13}, 
J.~Montgomery\textsuperscript{17}, 
Y.~Nakato\textsuperscript{9}, 
T.~Natoli\textsuperscript{2}, 
G.~I.~Noble\,\orcidlink{0000-0002-5254-243X}\textsuperscript{32,33}, 
Y.~Omori\textsuperscript{3,2}, 
S.~Padin\textsuperscript{2,3,34}, 
Z.~Pan\,\orcidlink{0000-0002-6164-9861}\textsuperscript{7,2,1}, 
P.~Paschos\textsuperscript{11}, 
K.~A.~Phadke\,\orcidlink{0000-0001-7946-557X}\textsuperscript{15,31}, 
A.~W.~Pollak\textsuperscript{12}, 
K.~Prabhu\textsuperscript{24}, 
W.~Quan\textsuperscript{7,1,2}, 
M.~Rahimi\textsuperscript{5}, 
C.~L.~Reichardt\,\orcidlink{0000-0003-2226-9169}\textsuperscript{5}, 
M.~Rouble\textsuperscript{17}, 
J.~E.~Ruhl\textsuperscript{29}, 
E.~Schiappucci\textsuperscript{5}, 
J.~A.~Sobrin\,\orcidlink{0000-0001-6155-5315}\textsuperscript{4,2}, 
A.~A.~Stark\textsuperscript{35}, 
J.~Stephen\textsuperscript{11}, 
C.~Tandoi\textsuperscript{15}, 
B.~Thorne\textsuperscript{24}, 
C.~Trendafilova\textsuperscript{31}, 
C.~Umilta\,\orcidlink{0000-0002-6805-6188}\textsuperscript{21}, 
J.~Veitch-Michaelis\textsuperscript{12}, 
J.~D.~Vieira\textsuperscript{15,21,31}, 
A.~Vitrier\textsuperscript{6}, 
Y.~Wan\textsuperscript{15,31}, 
N.~Whitehorn\,\orcidlink{0000-0002-3157-0407}\textsuperscript{23}, 
W.~L.~K.~Wu\,\orcidlink{0000-0001-5411-6920}\textsuperscript{8,10}, 
M.~R.~Young\textsuperscript{4,2}, 
K.~Zagorski\textsuperscript{12} 
and J.~A.~Zebrowski\textsuperscript{2,3,4}
}
\address{
\textsuperscript{1}Department of Physics, University of Chicago, 5640 South Ellis Avenue, Chicago, IL, 60637, USA\\
\textsuperscript{2}Kavli Institute for Cosmological Physics, University of Chicago, 5640 South Ellis Avenue, Chicago, IL, 60637, USA\\
\textsuperscript{3}Department of Astronomy and Astrophysics, University of Chicago, 5640 South Ellis Avenue, Chicago, IL, 60637, USA\\
\textsuperscript{4}Fermi National Accelerator Laboratory, MS209, P.O. Box 500, Batavia, IL, 60510, USA\\
\textsuperscript{5}School of Physics, University of Melbourne, Parkville, VIC 3010, Australia\\
\textsuperscript{6}Sorbonne Universit\'e, CNRS, UMR 7095, Institut d'Astrophysique de Paris, 98 bis bd Arago, 75014 Paris, France\\
\textsuperscript{7}High-Energy Physics Division, Argonne National Laboratory, 9700 South Cass Avenue., Lemont, IL, 60439, USA\\
\textsuperscript{8}Kavli Institute for Particle Astrophysics and Cosmology, Stanford University, 452 Lomita Mall, Stanford, CA, 94305, USA\\
\textsuperscript{9}Department of Physics, Stanford University, 382 Via Pueblo Mall, Stanford, CA, 94305, USA\\
\textsuperscript{10}SLAC National Accelerator Laboratory, 2575 Sand Hill Road, Menlo Park, CA, 94025, USA\\
\textsuperscript{11}Enrico Fermi Institute, University of Chicago, 5640 South Ellis Avenue, Chicago, IL, 60637, USA\\
\textsuperscript{12}University of Chicago, 5640 South Ellis Avenue, Chicago, IL, 60637, USA\\
\textsuperscript{13}Department of Physics, University of California, Berkeley, CA, 94720, USA\\
\textsuperscript{14}Universit\'e Paris-Saclay, Universit\'e Paris Cit\'e, CEA, CNRS, AIM, 91191, Gif-sur-Yvette, France\\
\textsuperscript{15}Department of Astronomy, University of Illinois Urbana-Champaign, 1002 West Green Street, Urbana, IL, 61801, USA\\
\textsuperscript{16}High Energy Accelerator Research Organization (KEK), Tsukuba, Ibaraki 305-0801, Japan\\
\textsuperscript{17}Department of Physics and McGill Space Institute, McGill University, 3600 Rue University, Montreal, Quebec H3A 2T8, Canada\\
\textsuperscript{18}Canadian Institute for Advanced Research, CIFAR Program in Gravity and the Extreme Universe, Toronto, ON, M5G 1Z8, Canada\\
\textsuperscript{19}Joseph Henry Laboratories of Physics, Jadwin Hall, Princeton University, Princeton, NJ 08544, USA\\
\textsuperscript{20}Department of Astrophysical and Planetary Sciences, University of Colorado, Boulder, CO, 80309, USA\\
\textsuperscript{21}Department of Physics, University of Illinois Urbana-Champaign, 1110 West Green Street, Urbana, IL, 61801, USA\\
\textsuperscript{22}Department of Physics and Astronomy, University of California, Los Angeles, CA, 90095, USA\\
\textsuperscript{23}Department of Physics and Astronomy, Michigan State University, East Lansing, MI 48824, USA\\
\textsuperscript{24}Department of Physics \& Astronomy, University of California, One Shields Avenue, Davis, CA 95616, USA\\
\textsuperscript{25}Department of Physics and Astronomy, Northwestern University, 633 Clark St, Evanston, IL, 60208, USA\\
\textsuperscript{26}CASA, Department of Astrophysical and Planetary Sciences, University of Colorado, Boulder, CO, 80309, USA \\
\textsuperscript{27}Department of Physics, University of Colorado, Boulder, CO, 80309, USA\\
\textsuperscript{28}Department of Physics, Korea Advanced Institute of Science and Technology (KAIST), 291 Daehak-ro, Yuseong-gu, Daejeon 34141, Republic of Korea\\
\textsuperscript{29}Department of Physics, Case Western Reserve University, Cleveland, OH, 44106, USA\\
\textsuperscript{30}Steward Observatory, University of Arizona, 933 North Cherry Avenue, Tucson, AZ 85721, USA\\
\textsuperscript{31}Center for AstroPhysical Surveys, National Center for Supercomputing Applications, Urbana, IL, 61801, USA\\
\textsuperscript{32}Dunlap Institute for Astronomy \& Astrophysics, University of Toronto, 50 St. George Street, Toronto, ON, M5S 3H4, Canada\\
\textsuperscript{33}David A. Dunlap Department of Astronomy \& Astrophysics, University of Toronto, 50 St. George Street, Toronto, ON, M5S 3H4, Canada\\
\textsuperscript{34}California Institute of Technology, 1200 East California Boulevard., Pasadena, CA, 91125, USA\\
\textsuperscript{35}Harvard-Smithsonian Center for Astrophysics, 60 Garden Street, Cambridge, MA, 02138, USA\\
\textsuperscript{*}pchichura@uchicago.edu\\
}

%% file: pointing.bbl
\begin{thebibliography}{29}
\newcommand{\enquote}[1]{``#1''}
\providecommand{\natexlab}[1]{#1}
\providecommand{\url}[1]{\texttt{#1}}
\providecommand{\urlprefix}{URL }
\expandafter\ifx\csname urlstyle\endcsname\relax
  \providecommand{\doi}[1]{doi:\discretionary{}{}{}#1}\else
  \providecommand{\doi}{doi:\discretionary{}{}{}\begingroup
  \urlstyle{rm}\Url}\fi

\bibitem[{{Antebi} \emph{et~al.}(1998){Antebi}, {Kan} \& {Rao}}]{antebi98}
{Antebi}, J., {Kan}, F.~W. \& {Rao}, R.~S. [1998]  \enquote{{Active segmented
  primary reflector and pointing accuracy of the Large Millimeter Telescope
  (LMT)},}  \emph{Advanced Technology MMW, Radio, and Terahertz Telescopes},
  ed. {Phillips}, T.~G., p. 220, \doi{10.1117/12.317356}.

\bibitem[{{Baars} \emph{et~al.}(1999){Baars}, {Martin}, {Mangum}, {McMullin} \&
  {Peters}}]{baars99}
{Baars}, J. W.~M., {Martin}, R.~N., {Mangum}, J.~G., {McMullin}, J.~P. \&
  {Peters}, W.~L. [1999]  \emph{Publications of the ASP} \textbf{111},  627,
  \doi{10.1086/316365}.

\bibitem[{{Carlstrom} \emph{et~al.}(2011){Carlstrom}, {Ade}, {Aird}
  \emph{et~al.}}]{carlstrom11}
{Carlstrom}, J.~E., {Ade}, P.~A.~R., {Aird}, K.~A. \emph{et~al.} [2011]
  \emph{Publications of the ASP} \textbf{123},  568, \doi{10.1086/659879}.

\bibitem[{{Chen} \& {Guestrin}(2016)}]{chen16}
{Chen}, T. \& {Guestrin}, C. [2016]  \enquote{{XGBoost}: A scalable tree
  boosting system,}  \emph{Proceedings of the 22nd ACM SIGKDD International
  Conference on Knowledge Discovery and Data Mining}, KDD '16 (ACM, New York,
  NY, USA), ISBN 978-1-4503-4232-2, p. 785, \doi{10.1145/2939672.2939785},
  \urlprefix\url{http://doi.acm.org/10.1145/2939672.2939785}.

\bibitem[{{Doeleman} \emph{et~al.}(2009){Doeleman}, {Agol}, {Backer}
  \emph{et~al.}}]{doeleman10}
{Doeleman}, S., {Agol}, E., {Backer}, D. \emph{et~al.} [2009]
  \enquote{{Imaging an Event Horizon: submm-VLBI of a Super Massive Black
  Hole},}  \emph{astro2010: The Astronomy and Astrophysics Decadal Survey}.

\bibitem[{{Dong} \emph{et~al.}(2018){Dong}, {Fu}, {Liu} \& {Shen}}]{dong18}
{Dong}, J., {Fu}, L., {Liu}, Q. \& {Shen}, Z. [2018]  \emph{Experimental
  Astronomy} \textbf{45},  397, \doi{10.1007/s10686-018-9592-3}.

\bibitem[{{Event Horizon Telescope Collaboration}
  \emph{et~al.}(2019{\natexlab{a}}){Event Horizon Telescope Collaboration},
  {Akiyama}, {Alberdi}, {Alef} \emph{et~al.}}]{eht19a}
{Event Horizon Telescope Collaboration}, {Akiyama}, K., {Alberdi}, A., {Alef},
  W. \emph{et~al.} [2019{\natexlab{a}}]  \emph{Astrophysical Journal, Letters}
  \textbf{875}, L1, \doi{10.3847/2041-8213/ab0ec7}.

\bibitem[{{Event Horizon Telescope Collaboration}
  \emph{et~al.}(2019{\natexlab{b}}){Event Horizon Telescope Collaboration},
  {Akiyama}, {Alberdi}, {Alef} \emph{et~al.}}]{eht19b}
{Event Horizon Telescope Collaboration}, {Akiyama}, K., {Alberdi}, A., {Alef},
  W. \emph{et~al.} [2019{\natexlab{b}}]  \emph{Astrophysical Journal, Letters}
  \textbf{875}, L2, \doi{10.3847/2041-8213/ab0c96}.

\bibitem[{{Event Horizon Telescope Collaboration} \emph{et~al.}(2022){Event
  Horizon Telescope Collaboration}, {Akiyama}, {Alberdi}, {Alef}
  \emph{et~al.}}]{eht22a}
{Event Horizon Telescope Collaboration}, {Akiyama}, K., {Alberdi}, A., {Alef},
  W. \emph{et~al.} [2022]  \emph{Astrophysical Journal, Letters} \textbf{930},
  L12, \doi{10.3847/2041-8213/ac6674}.

\bibitem[{{Gawronski} \& {Souccar}(2005)}]{gawronski05}
{Gawronski}, W. \& {Souccar}, K. [2005]  \emph{IEEE Antennas and Propagation
  Magazine} \textbf{47},  41, \doi{10.1109/MAP.2005.1608713}.

\bibitem[{{Greve} \& {Bremer}(2010)}]{greve10}
{Greve}, A. \& {Bremer}, M. [2010]  \emph{{Thermal Design and Thermal Behaviour
  of Radio Telescopes and their Enclosures}}, Vol.~364,
  \doi{10.1007/978-3-642-03866-2}.

\bibitem[{{Greve} \emph{et~al.}(1996){Greve}, {Panis} \& {Thum}}]{greve96}
{Greve}, A., {Panis}, J.~F. \& {Thum}, C. [1996]  \emph{Astronomy and
  Astrophysics, Supplement} \textbf{115},  379.

\bibitem[{{Hou} \emph{et~al.}(2023){Hou}, {Hu}, {Du} \emph{et~al.}}]{hou23}
{Hou}, X., {Hu}, Y., {Du}, F. \emph{et~al.} [2023]  \emph{Astronomy and
  Computing} \textbf{43}, 100710, \doi{10.1016/j.ascom.2023.100710}.

\bibitem[{{Ivison} \emph{et~al.}(2007){Ivison}, {Greve}, {Dunlop}
  \emph{et~al.}}]{ivison07}
{Ivison}, R.~J., {Greve}, T.~R., {Dunlop}, J.~S. \emph{et~al.} [2007]
  \emph{Monthly Notices of the RAS} \textbf{380},  199,
  \doi{10.1111/j.1365-2966.2007.12044.x}.

\bibitem[{{Kim} \emph{et~al.}(2018{\natexlab{a}}){Kim}, {Marrone}, {Beaudoin}
  \emph{et~al.}}]{kim18b}
{Kim}, J., {Marrone}, D.~P., {Beaudoin}, C. \emph{et~al.} [2018{\natexlab{a}}]
  \emph{Proceedings of the SPIE} \textbf{107082S}.

\bibitem[{{Kim} \emph{et~al.}(2018{\natexlab{b}}){Kim}, {Marrone}, {Roy}
  \emph{et~al.}}]{kim18a}
{Kim}, J., {Marrone}, D.~P., {Roy}, A.~L. \emph{et~al.} [2018{\natexlab{b}}]
  \emph{Astrophysical Journal} \textbf{861}, 129,
  \doi{10.3847/1538-4357/aac7c6}.

\bibitem[{{Lazzara} \emph{et~al.}(2012){Lazzara}, {Keller}, {Markle} \&
  {Gallagher}}]{lazzara12}
{Lazzara}, M.~A., {Keller}, L.~M., {Markle}, T. \& {Gallagher}, J. [2012]
  \emph{Atmospheric Research} \textbf{118},  240,
  \doi{10.1016/j.atmosres.2012.06.027}.

\bibitem[{{Mittag} \emph{et~al.}(2008){Mittag}, {Hempelmann}, {Gonzalez-Perez}
  \& {Schmitt}}]{mittag08}
{Mittag}, M., {Hempelmann}, A., {Gonzalez-Perez}, J.~N. \& {Schmitt},
  J.~H.~M.~M. [2008]  \emph{Publications of the ASP} \textbf{120},  425,
  \doi{10.1086/533478}.

\bibitem[{{Nyheim} \emph{et~al.}(2024){Nyheim}, {Riemer-S{\o}rensen}, {Parra}
  \& {Cicone}}]{nyheim24}
{Nyheim}, B., {Riemer-S{\o}rensen}, S., {Parra}, R. \& {Cicone}, C. [2024]
  \emph{arXiv e-prints} , arXiv:2402.08589\doi{10.48550/arXiv.2402.08589}.

\bibitem[{{Padin} \emph{et~al.}(2002){Padin}, {Shepherd}, {Cartwright}
  \emph{et~al.}}]{padin02}
{Padin}, S., {Shepherd}, M.~C., {Cartwright}, J.~K. \emph{et~al.} [2002]
  \emph{Publications of the ASP} \textbf{114},  83, \doi{10.1086/324786}.

\bibitem[{{Prabhu} \emph{et~al.}(2024){Prabhu}, {Raghunathan}, {Millea}
  \emph{et~al.}}]{prabhu24}
{Prabhu}, K., {Raghunathan}, S., {Millea}, M. \emph{et~al.} [2024]
  \emph{Astrophysical Journal} \textbf{973}, 4, \doi{10.3847/1538-4357/ad5ff1}.

\bibitem[{{Shwartz-Ziv} \& {Armon}(2022)}]{shwartz22}
{Shwartz-Ziv}, R. \& {Armon}, A. [2022]  \emph{Information Fusion} \textbf{81},
   84.

\bibitem[{{Sobrin} \emph{et~al.}(2022){Sobrin}, {Anderson}, {Bender}
  \emph{et~al.}}]{sobrin22}
{Sobrin}, J.~A., {Anderson}, A.~J., {Bender}, A.~N. \emph{et~al.} [2022]
  \emph{Astrophysical Journal, Supplement} \textbf{258}, 42,
  \doi{10.3847/1538-4365/ac374f}.

\bibitem[{{Story} \emph{et~al.}(2012){Story}, {Leitch}, {Ade}
  \emph{et~al.}}]{story12}
{Story}, K., {Leitch}, E., {Ade}, P. \emph{et~al.} [2012]  \enquote{{South Pole
  Telescope software systems: control, monitoring, and data acquisition},}
  \emph{Software and Cyberinfrastructure for Astronomy II}, eds. {Radziwill},
  N.~M. \& {Chiozzi}, G., p. 84510T, \doi{10.1117/12.925808}.

\bibitem[{{Stumpff}(1972)}]{stumpff72}
{Stumpff}, P. [1972]  \emph{Klein-Heibacher Berichte} \textbf{15},  431.

\bibitem[{{Ulich}(1981)}]{ulich81}
{Ulich}, B.~L. [1981]  \emph{International Journal of Infrared and Millimeter
  Waves} \textbf{2},  293, \doi{10.1007/BF01007036}.

\bibitem[{{von Hoerner} \& {Wong}(1975)}]{vonhoerner75}
{von Hoerner}, S. \& {Wong}, W.~Y. [1975]  \emph{IEEE Transactions on Antennas
  and Propagation} \textbf{23},  689, \doi{10.1109/TAP.1975.1141163}.

\bibitem[{{Wallace}(1994)}]{wallace94}
{Wallace}, P.~T. [1994]  \enquote{{The SLALIB Library},}  \emph{Astronomical
  Data Analysis Software and Systems III}, eds. {Crabtree}, D.~R., {Hanisch},
  R.~J. \& {Barnes}, J., p. 481.

\bibitem[{{White} \emph{et~al.}(2022){White}, {Ghigo}, {Prestage}
  \emph{et~al.}}]{white22}
{White}, E., {Ghigo}, F.~D., {Prestage}, R.~M. \emph{et~al.} [2022]
  \emph{Astronomy and Astrophysics} \textbf{659}, A113,
  \doi{10.1051/0004-6361/202141936}.

\end{thebibliography}
